\documentclass[a4paper,11pt]{JHEP3}
\usepackage{amssymb}
\usepackage{amsmath}
\usepackage[dvips]{graphicx}

\newcommand{\ket}[1]{|\,#1\,\rangle}

\def\u{u}
\title{Towards a non-abelian electric-magnetic symmetry: the skeleton group }
\author{L.~Kampmeijer,$^{a}$ F.\,A.~Bais,$^{a,b}$ B.\,J.~Schroers\,$^c$ and J.\,K.~Slingerland\,$^{d,e}$\\
{}$^a$ Institute for Theoretical Physics, University of Amsterdam, Valckenierstraat 65,\\ 1018 XE Amsterdam, The Netherlands\\
{}$^b$ Santa Fe Institute, 1120 Canyon Rd, Santa Fe, NM 87501, USA\\
{}$^c$ Department of Mathematics and Maxwell Institute for Mathematical Sciences, Heriot-Watt University, Edinburgh EH14 4AS, United Kingdom \\
{}$^d$ Dublin Institute for Advanced Studies, School for Theoretical Physics,\\
10 Burlington Rd, Dublin, Ireland \\
{}$^e$ Department of Mathematical Physics, National University of Ireland,  Maynooth,\\
Co. Kildare, Ireland\\
E-mail: \email{leo.kampmeijer@uva.nl}, \email{f.a.bais@uva.nl}, \email{bernd@ma.hw.ac.uk}, \email{slingerland@stp.dias.ie}
}
\abstract{%
We propose an electric-magnetic symmetry group in non-abelian gauge theory, which we call the skeleton group.  We work in the context of  non-abelian unbroken gauge symmetry,  and provide evidence for our proposal by relating the representation theory of the skeleton group to  the labelling and fusion rules  of charge sectors. We show that the labels of electric, magnetic and dyonic sectors  in non-abelian Yang-Mills theory  can be interpreted  in terms of irreducible representations of the skeleton group.  Decomposing  tensor products of these representations thus gives a set of fusion rules  which contain information about the full fusion rules of these charge sectors. We demonstrate consistency of the skeleton's fusion rules with the known fusion rules of the purely electric and purely magnetic magnetic sectors,  and extract new predictions for the fusion rules of dyonic sectors in particular cases. We also implement S-duality  and show that the fusion  rules  obtained from the skeleton group commute with S-duality.}
\preprint{
ITFA-2008-47\\
EMPG-08-19\\
DIAS-STP-08-14
}
\keywords{Monopoles, Gauge Symmetry, Electric-Magnetic  Symmetry,  Duality in Gauge Field Theories}
\begin{document}
\section{Introduction}                                                            %
In this paper we try to determine the electric-magnetic symmetry in a non-abelian gauge theory. This task may be formulated in many ways, varying in physical content and mathematical sophistication. Our main goal is to find a consistent large distance description of the electric, magnetic and dyonic degrees of freedom. More specifically, we would  like to uncover a hidden algebraic structure which governs the labelling  and the fusion rules of the physical sectors in general gauge theories and which is compatible with S-duality.  While there are compelling physical arguments for the existence of such a structure, we are aware that its existence is not guaranteed, and that there is, at present, no method for determining it deductively. Our strategy is therefore to propose a solution  and then to check its consistency with known facts about the charge sectors in non-abelian gauge theories.  One important motivation of this paper is the prospect of  using a better understanding of electric-magnetic symmetry for studying the    phases of non-abelian gauge theory.  We do not carry out such a study here, but at the end of the paper we sketch how it might proceed.

The standard literature on electric-magnetic symmetry  is based on the dual symmetry proposed by Goddard, Nuyts and Olive \cite{Goddard:1976qe}. Following earlier work of Englert and Windey on the generalised Dirac quantisation condition \cite{Englert:1976ng} they showed that the magnetic charges of monopoles in a theory with gauge group $G$ take values in the weight lattice of the dual gauge group  $G^*$, now known as the GNO or Langlands dual group. Based on this fact they came up with a bold yet attractive conjecture: monopoles transform in representations of the dual group.

Considering the fact that the Bogomolny Prasad Sommerfeld (BPS) mass formula for dyons \cite{Bogomolny:1975de,Prasad:1975kr} is invariant under the interchange of electric and magnetic quantum numbers if the coupling constant is inverted as well, Montonen and Olive  extended the GNO conjecture. Their proposal was that the strong coupling regime of some suitable quantum field theory is described by a weakly
coupled theory with a similar Lagrangian but with the gauge group replaced by the GNO dual group and the coupling constant inverted \cite{Montonen:1977sn}; in other words, the dual gauge symmetry is manifestly realised in the strongly coupled phase of the theory.

The non-abelian version of the Montonen-Olive conjecture has been proven by Kapustin and Witten \cite{Kapustin:2006pk} for a twisted $\mathcal N=4$ supersymmetric Yang-Mills theory. Using the identification of singular monopoles with 't Hooft operators and computing the operator product expansion (OPE) for the latter, they showed that 
the fusion rules of purely magnetic monopoles  are identical to the fusion rules of the dual gauge group. It was shown in \cite{Kampmeijer} that, in ordinary $\mathcal N=4$ Yang-Mills theory, the classical fusion rules of monopoles, obtained from patching together monopole solutions to the classical field equations, are also consistent with the non-abelian Montonen-Olive conjecture.

A stronger version of the GNO conjecture is that a gauge theory has a hidden electric-magnetic symmetry of the type $G\times G^*$. The problem with this proposal is that the dyonic sectors do not respect this symmetry in phases where one has a residual non-abelian gauge symmetry. In such phases it may be that in a given magnetic sector there is an obstruction to the implemention of the full electric group. In a monopole background the  global electric symmetry is restricted to the centraliser in $G$ of the magnetic charge \cite{Abouelsaood:1982dz,Abouelsaood:1983gw, Nelson:1983bu,Balachandran:1983fg, Horvathy:1984yg, Horvathy:1985bp}. Dyonic charge sectors are thus not labelled by a $G\times G^*$ representation but instead, up to gauge transformations, by a magnetic charge and an electric centraliser representation \cite{Bais:1997qy}. This interplay of electric and magnetic degrees of freedom is not captured by the $G \times G^*$ structure. It is therefore a challenge to find a different  algebraic structure which underlies  the complicated pattern of the different electric-magnetic sectors in a non-abelian phase. 

We thus arrive at a list of requirements for the sought-after  algebraic structure.  It  would  have to reproduce  the complete set of fusion rules for all the different sectors, which is not known at present, and in particular would have to combine  within one framework the different centraliser groups that may occur  for dyons. It also has to be consistent with the labelling of purely electric sectors by irreducible  representations of the  full electric gauge group $G$ and with their fusion, as described by the tensor product decomposition of  $G$ representations. Similarly one should require that  purely magnetic sectors (at least for the twisted $\mathcal N=4$  theory considered by Kapustin and Witten in \cite{Kapustin:2006pk}) be labelled by irreducible representations of the magnetic gauge group $G^*$, with fusion rules given by  the $G^*$ representation theory.

Recently, Kapustin  found a  labelling of  all charge sectors   -  electric, magnetic and dyonic - in terms of  the set $(\Lambda\times \Lambda^*)/\mathcal W$, where $\mathcal W$ is the Weyl group (which isomorphic for $G$ and $G^*$) and $\Lambda$ and $\Lambda^*$ are the weight lattices of, respectively, $G$ and $G^*$ \cite{Kapustin:2005py}. 
In this paper we start with Kapustin's labelling  and,  generalising an earlier proposal by two of the authors \cite{Schroers:1998pg}, we introduce the \emph{skeleton group} $S$ as a candidate for the electric-magnetic symmetry group in a  non-abelian gauge theory. The skeleton group is,  in general,  a non-abelian group which  manifestly includes at least part of the 
 non-abelian electric and magnetic symmetry.  It has a simple definition  \eqref{newsdef}
  as a certain subgroup of $G\times G^*$, which only uses data naturally associated to $G$ and $G^*$  and which readers familiar with standard Lie algebra notation are invited to inspect at this stage.   Irreducible representations of $S$ exist for each charge sector and therefore  the representation theory of $S$ provides us with a consistent set of fusion rules for  all  charge sectors, including the dyonic ones.

The skeleton group does not completely fulfill our original objective. It has roughly the product structure  $\mathcal{W} \ltimes (T\times T^*)$  
where  $T$ and $T^*$  are the maximal tori  of $G$ and $G^*$,  and therefore   contains neither the full electric gauge group $G$ nor the magnetic group $G^*$.  As  a result,  its representation theory does not  reproduce  the representation theories of either $G$ or $G^*$. 
However, it is consistent with all the results of this paper to conjecture that the skeleton group is a subgroup of the full symmetry object governing the theory which can be realized in every electric-magnetic charge sector. In particular, since the skeleton group is a subgroup of $G\times G^*$  both the purely electric and the purely magnetic  representation theory of the skeleton group is consistent with the representation theories of $G$  and $G^*$ in a sense that we will explain. 

One should expect the dyonic sectors and fusion rules to be robust and in particular independent of the dynamical details of the particular model.  In this paper we therefore do not consider specific models. Nonetheless, our results must be consistent with what is known, for example, about S-duality of $\mathcal N=4$ super Yang-Mills theories.  The skeleton approach indeed allows for an explicit implementation of S-duality on its representation content, and we are able to show that
the fusion  rules  obtained from the skeleton group commute with S-duality. Thus the skeleton proposal leads to fusion rules which are invariant under S-duality.

The outline of the paper is as follows. After introducing our conventions and notation in section  \ref{sect:lieconventions}, we explain, in section \ref{sect:chargesectors}, the equivalence between the labelling of dyonic charge sectors involving centraliser representations and the labelling introduced by Kapustin \cite{Kapustin:2005py}. In section \ref{sect:skelgroup} we introduce the skeleton group as a candidate for a unified electric-magnetic symmetry group in Yang-Mills theory. A substantial part of this section is taken up by a detailed exposition of various aspects of the skeleton group which are needed in subsequent sections. In section \ref{sect:repsskel} we provide evidence for the relevance of the skeleton group by relating the representation theory of the skeleton group to the labelling and fusion rules of charge sectors. In particular, we show that the labels of electric, magnetic and dyonic sectors in a non-abelian Yang-Mills theory can be interpreted  in terms of irreducible representations of the skeleton group.  Decomposing  tensor products of irreducible representations of the skeleton group thus gives candidate fusion rules for these charge sectors. We demonstrate consistency of these fusion rules with the known fusion rules of the purely electric or magnetic sectors, and extract new predictions for the fusion rules of dyonic sectors in particular cases.  Section \ref{sect:sduality} contains a brief review of S-duality and its action on dyonic charge sectors. We define an action of S-duality on the irreducible representations of the skeleton group, and show that the fusion rules predicted by the skeleton group are invariant under this action. The final section \ref{outlook} contains an outlook onto possible uses of the skeleton group in studying phases of non-abelian gauge theories. 
\section{Lie algebra conventions}
\label{sect:lieconventions}
We briefly summarise some facts and conventions that we shall use in the subsequent sections regarding Lie algebras and Lie groups. Additional background material can be found, e.g., in \cite{Fuchs:1997jv}. We consider a semi-simple Lie algebra $\mathfrak g$ of rank $r$ and use $\mathfrak t$  to denote a fixed Cartan subalgebra (CSA).  The requirement of semi-simplicity is not strictly necessary for most of what we say in this paper  but it  allows us to make use of a Killing form on the Lie algebra from the outset,  and to  use a  unified notation for the CSA and its dual. It is worth emphasising that the  Killing form  is only indispensable for the discussion of S-duality  in section \ref{sect:sduality}; a related discussion   that avoids the use of the Killing form as far as possible can be found in    \cite{Kapustin:2005py}.
 
  We write $H$ for  an arbitrary element in $\mathfrak t$; for definiteness we shall often work with a basis $\{H_1,\ldots,H_r\}$ of $\mathfrak t$ which is orthonormal with respect to the Killing form $\langle.\,,.\rangle$ restricted to $\mathfrak t$. 
Then the Lie brackets of $\mathfrak g$  take the following form in the Cartan-Weyl basis of $\mathfrak g$:
\begin{equation}
\label{eqn:CartanWeyl}
[H_i, E_\alpha] = \alpha_i E_\alpha \qquad [E_\alpha, E_{-\alpha}]= \frac{2  \alpha_i  H_i}{\alpha_i \alpha_i} \equiv \frac{2  \alpha \cdot H}{\alpha^2}.
\end{equation}
The $r$-dimensional vectors $\alpha = (\alpha_i)_{i=1,\dots,r}$ are the root vectors  of $\mathfrak g$ relative to the basis $\{H_i \}_{i=1,\dots,r}$.  We use the dot notation to denote the contraction between the indices, with repeated indices automatically summed over. Also note that $\alpha^2\equiv\alpha\cdot\alpha$. 

Each root $\alpha$ can naturally be interpreted as an element in $\mathfrak t^*$, i.e.,  as a linear map  which assigns to $H\in \mathfrak  t  $ the (generally complex)  number 
 $\alpha(H)$ defined via
\begin{equation}
\label{rootint}
[H,E_\alpha]=\alpha(H) E_\alpha.
\end{equation}
The interpretation of roots as elements of $\mathfrak t^*$ is  fundamental and independent of the
inner product $\langle.,.\rangle$ on $\mathfrak t$. Comparing with \ref{eqn:CartanWeyl} we see that, with our conventions, 
\begin{equation}
\alpha(H_i)=\alpha_i, 
\end{equation}
and thus recover the root vector $(\alpha_i)_{i=1,\dots,r}$ relative to the basis $\{H_i\}_{i=1,\dots,r}$.   The relation between  $\alpha \in \mathfrak t^*$ and an $r$-component vector  depends on both the inner product $\langle.,.\rangle$  and the choice of basis  $\{H_i\}$ of $\mathfrak t$. Nonetheless, we will use the same notation for both in this paper. It will be clear from the context if we are thinking of it as a map $\mathfrak t\rightarrow \mathbb C$ or as an $r$-component vector.

The equation \eqref{rootint} shows that roots are eigenvalues of elements $H \in \mathfrak t$ in the adjoint representation. More generally, eigenvalues of  elements $H\in \mathfrak t$  in an  arbitrary repesentation of  $\mathfrak g$  are called weights; like roots they are naturally elements of $\mathfrak t^*$.

Instead of the basis $\{H_i\}_{i=1,\dots,r}$ for $\mathfrak t$ one can choose a basis  associated to  simple roots (which span the root space with integer coefficients which are either all positive or all negative) via 
\begin{equation}
\label{halphadef}
H_\alpha = 2\alpha^*\cdot  H,
\end{equation}
where $\alpha^* = \alpha/ \alpha^2$.
The elements $H_\alpha$ are called coroots; they satisfy
\begin{equation}
[H_\alpha, E_\beta] = 2 \alpha^*\cdot \beta  \;E_{\beta} \qquad [E_\alpha, E_{-\alpha}]= H_\alpha.
\end{equation}

The lattices generated by  roots and coroots play  a fundamental role in Lie algebra theory. 
The roots span the \emph{root lattice} $\Lambda_r \subset \mathfrak t^*$ and  the coroots span the \emph{coroot lattice} $\Lambda_{cr}\subset \mathfrak t$.   The dual lattice of the coroot lattice is called the \emph{weight lattice} $\Lambda_w\subset \mathfrak t^*$  of $\mathfrak g$ and is generated by the fundamental weights of $\mathfrak g$. The dual lattice of the root lattice is  called the \emph{magnetic weight lattice} $\Lambda_{mw}\subset \mathfrak t$. 

So far our review has only been concerned with the Lie algebra $\mathfrak g$. Consider now a connected Lie group $G$ with Lie algebra $\mathfrak g$. The group $G$ and all other  Lie groups appearing in this paper are assumed to be compact. Any representation of $G$ gives rise to a representation of $\mathfrak g$ and thus to an associated weight lattice. 
The weight lattice $\Lambda(G)$ of a Lie group $G$ with Lie algebra $\mathfrak g$  satisfies 
\begin{equation}
\Lambda_r \subset \Lambda(G) \subset \Lambda_w,
\end{equation}
while the dual weight lattice $\Lambda^*(G)$ satisfies 
\begin{equation}
\Lambda_{cr}\subset \Lambda^*(G) \subset \Lambda_{mw}.
\end{equation}
$\Lambda^*(G)$ can be identified with the weight lattice $\Lambda(G^*)$ of the GNO dual group $G^*$ \cite{Goddard:1976qe}. The roots of $G^*$ correspond to the coroots of $G$ while the fundamental weights of $G^*$ span $\Lambda_{mw}$. These relations are summarised in table  1. This table also summarises other notational conventions that will be used in subsequent sections as well as various relations that will be discussed below.
\begin{table}[!ht]
\begin{small}
\[
\begin{array}{|ccccccccc|}
\hline
&&&&&&&&\\
&&\text{Weyl group}\;\;\;\makebox[17mm]{\hspace*{3mm}$ \mathcal{W}^*\;=\;\mathcal{W}\;\simeq\;$} & \widetilde{W}^*/\widetilde{D}^* &\simeq & W^*/D^* &\simeq &\overline{W}^*/\overline{D}^*&\\
&&\uparrow &  & \uparrow & & \uparrow  &&\\
&&\text{Lifted Weyl group}&\widetilde{W}^*& \leftarrow& W^* & \leftarrow &  \overline{W}^*&\\
&&&\cap &  & \cap & & \cap  &\\
&{\bf magnetic}&\text{Dual gauge  group} &\widetilde{G}^* = \overline{G}^*/Z^* & \leftarrow& G^* & \leftarrow &\overline{G}^* &\\
&&&\cup&&\cup&&\cup&\\
&&\text{Dual torus} &\widetilde{T}^*= \mathfrak{t}_{\mathbb R}^*/\Lambda_{w} & \leftarrow& T^* = \mathfrak{t}_{\mathbb R}^*/\Lambda & \leftarrow &\overline{T}^*  =  \mathfrak{t}_{\mathbb R}^*/\Lambda_{r} &\\
&&& &  &  & &  &\\
&&\text{Dual weight lattice} &\widetilde{\Lambda}^*  = \Lambda_{cr}  & \subset & \Lambda^* & \subset & \overline{\Lambda}^* = \Lambda_{mw} &\\
&&& & & & &&\\
\hline
&&&&&&& &\\
&&\text{Weight lattice}&\widetilde{\Lambda}=\Lambda_w & \supset & \Lambda & \supset & \overline{\Lambda} =\Lambda_{r} & \\
&&&&&&&&\\
&&\text{ Maximal  torus}&\widetilde{T}= \mathfrak{t}_{\mathbb R} /\Lambda_{cr} & \rightarrow& T = \mathfrak{t}_{\mathbb R}/\Lambda^* & \rightarrow &\overline{T}  =  \mathfrak{t}_{\mathbb R}/\Lambda_{mw} &\\
&&&\cap &  & \cap &  & \cap &\\
&{\bf  electric}&\text{Gauge group} &\widetilde{G}& \rightarrow& G & \rightarrow &\overline{G} = \widetilde{G}/Z &\\
&&&\cup &  & \cup & & \cup  &\\
&&\text{Lifted Weyl group}&\widetilde{W} & \rightarrow& W  & \rightarrow &  \overline{W}  &\\
&&&\downarrow &  & \downarrow & & \downarrow  &\\
&&\text{Weyl group } \makebox[1mm]{\hspace*{15mm}$\mathcal{W}\;\;\simeq$} & \widetilde{W}/\widetilde{D} &\simeq & W/D &\simeq &\overline{W}/\overline{D}&\\
&&&&&&&&\\
\hline 
\end{array}
\]
\caption{Notational conventions and relations regarding Lie algebras, Lie groups and Weyl groups.}
\end{small}
\end{table}

\vfill \pagebreak
\section{Charge sectors of non-abelian gauge  theories}
\label{sect:chargesectors}
One of the key features of the skeleton group is that it reproduces the dyonic charge sectors of a Yang-Mills theory. To appreciate this one needs some basic understanding of the electric and magnetic charge lattices and the set of dyonic charge sectors.

\subsection{Electric charge lattices}
To define the electric content of a gauge theory one starts by choosing an appropriate electric charge lattice $\Lambda$. Choosing an electric charge lattice corresponds to choosing a gauge group $G$ such that $\Lambda$ equals the weight lattice $\Lambda(G)$ of $G$.  The electric charge lattice $\Lambda$  can vary from the root lattice $\Lambda_r$ to the  weight lattice $\Lambda_w$ of $\mathfrak{g}$. This corresponds to the fact that for a fixed Lie algebra $\mathfrak g$ one can vary the Lie group $G$  from $\overline{G}$ all the way to $\widetilde{G}$, where $\widetilde {G}$ is the universal covering group of $G$ and $\overline{G}$ is the so-called adjoint group, which is the covering group divided by the centre $Z(\widetilde G)$. Note that the possible electric gauge groups are not related as subgroups but rather by taking quotients.

\subsection{Magnetic charge lattices}
Once the electric group $G$ is chosen one is free to choose the magnetic spectrum as long as the generalised Dirac quantisation condition \cite{Goddard:1976qe,Englert:1976ng} is respected.
The original treatment of that  condition in \cite{Goddard:1976qe,Englert:1976ng} makes use of 
the identification  between $\mathfrak t^*$ and  $\mathfrak t$ via the Killling form and uses a basis  in order to describe both  magnetic   and electric charges as $r$-component vectors. We have followed  that path in the current paper, and also in  \cite{Kampmeijer} where we give a review  using the same conventions as  in the current paper. It is worth emphasising, however, that the Dirac condition only requires the natural duality between 
$\mathfrak t$ (magnetic charges) and $\mathfrak t^*$
(electric charges) for its formulation, as stressed in \cite{Kapustin:2005py}. With the right normalisation it merely states that magnetic and electric charges lie on mutually dual lattices.
Thus, for a given electric group $G$ (and hence given electric charges) 
 the Dirac condition forces magnetic charges to lie on a lattice $\Lambda^*\subset  \mathfrak t$.
In fact,   one may also choose to  restrict magnetic charges to a sublattice:
  in analogy with the electric side, the choice of the magnetic charge lattice corresponds to fixing the magnetic group $G^*$ by demanding that its  weight lattice $\Lambda(G^*)$ equals $\Lambda^*$. 
  
Even though $G$  and $G^*$ do not uniquely determine each other, the generalised quantisation condition does put restrictions on the pair $(G,G^*)$. First of all, the roots of $G^*$ correspond to the coroots of $G$. Hence, the Lie algebra $\mathfrak{g}$ of $G$ uniquely fixes the Lie algebra $\mathfrak{g}^*$ of $G^*$ and vice versa. The universal covering groups $\widetilde G$ and $\overline {G}^*$ are therefore also uniquely related. Moreover, once $G$ is fixed, the Dirac quantisation condition tells us that the set of magnetic charges $\Lambda^*$ must be a subset of  $\Lambda^*(G)\subset\Lambda_{mw}$. Note that $\Lambda_{mw}$ is precisely the weight lattice of the universal covering group $\overline{G}^*$ of $G^*$. Taking $\Lambda^*$ equal to $\Lambda^*(G)$ amounts to choosing $G^*$ to be the GNO dual group of $G$. We thus see that, once $G$ is fixed,  $G^*$ can vary between the adjoint group $\widetilde{G}^*$ and the GNO dual group of $G$. Analogously, if $G^*$ is fixed $G$ can vary between the GNO dual of $G^*$  and the adjoint group $\overline G$ without violating the generalised Dirac quantisation condition.\\
\\
Unless stated otherwise we shall assume that all charges allowed by the Dirac quantisation condition occur and take $G$ and $G^*$ to be their respective GNO duals. Note that if the fields present in the Lagrangian are only adjoint fields and one only wants to consider smooth monopoles it is natural to restrict $G$ and $G^*$ to be adjoint groups.\\ 
%
\subsection{Dyonic charge sectors}
\label{subsect:dyonicsectors}
It was observed in \cite{Abouelsaood:1982dz,Abouelsaood:1983gw, Nelson:1983bu,Balachandran:1983fg, Horvathy:1984yg, Horvathy:1985bp} that in a monopole background the  global gauge symmetry is restricted to the centraliser $C_g$ of the magnetic charge $g$. This implies that the charges of dyons are given by a pair $(R_\lambda,g )$ where $g$ is the usual magnetic charge corresponding to an element in the Lie algebra of $G$ and $R_\lambda$ is an irreducible representation of $C_g\subset G$. It is explained in \cite{Kapustin:2005py} how these dyonic sectors can be relabelled in a convenient way. We shall give a brief review.

Since the magnetic charge is an element of the Lie algebra one can effectively view $C_g$ as the residual gauge group that arises from adjoint symmetry breaking where the Lie algebra valued Higgs VEV  is replaced by the magnetic charge. The Lie algebra of $\mathfrak g_g$ of  $C_g$ is easily determined. One can choose a gauge where the magnetic charge lies in a chosen CSA of $G$. 
Note that this does not fix $g$ uniquely since the intersection of its gauge orbit and the CSA corresponds to a complete Weyl orbit.
Now since the generators  of the CSA commute one immediately finds that the complete CSA of $G$ is contained in the Lie algebra of $C_g$. The remaining basis elements of $\mathfrak g_g$ are given by $E_\alpha$  with $\alpha(g)=0$, so that, in terms of the Killing metric, $\alpha$ is  perpendicular to $g$. This follows from the fact that $[E_\alpha,g]= \alpha(g) E_\alpha$. We thus see that the weight lattice of $C_g$ is identical to the weight lattice of $G$, whereas the roots of $C_g$ are a subset of the roots of $G$. Consequently  the Weyl group $\mathcal W_g$ of $C_g$ is the subgroup in the Weyl group $\mathcal W$ of $G$ generated by the reflections in the hyperplanes perpendicular to the roots of $C_g$.
An irreducible representation $R_\lambda$ of $C_g$ is uniquely labelled by a highest weight $\lambda$ of $C_g$ or, equivalently, a  $\mathcal W_g$ orbit $[\lambda]$ in the weight lattice of $C_g$, which is in particular a $\mathcal W_g$ orbit in the weight lattice of $G$. Remembering that $g$ itself is only fixed up to Weyl transformations, and using $C_g\simeq C_{w(g)}$ for all $w\in \mathcal W$ we find that a dyonic charge sector  is labelled  by an equivalence class $[\lambda, g]$  under the diagonal action of $\mathcal W$: every equivalence class   automatically includes a full Weyl orbit of $g$ and a $\mathcal W_g$ orbit of $\lambda$.

One of the goals of this paper is to find the fusion rules of dyons. We have explained that dyons are classified by an equivalence class of pairs $(\lambda, g)\in \Lambda(G)\times \Lambda(G^*)$ under the action of $\mathcal W$. By fusion rules we mean a set of rules of the form:
\begin{equation}
\label{eqn:fusionrules}
(R_{\lambda_1} , g_1) \otimes (R_{\lambda_2} ,g_2) = \bigoplus_{[\lambda, g]} N_{\lambda_1 ,\lambda_2 , g_1, g_2}^{\lambda,g} (R_{\lambda}, g),
\end{equation}
where the coefficients $N^{\lambda,g}_{\lambda_1 ,\lambda_2, g_1, g_2}$ are non-negative integers. These integers are non-vanishing only for a finite number of terms. One may also expect the product in equation \eqref{eqn:fusionrules} to be commutative and associative. Finally one would like the fusion rules  that follow from the representation ring of  $G$ and $G^*$  to be respected for the purely electric and the purely magnetic cases.  
\section{The skeleton group}
\label{sect:skelgroup}
In order to motivate our definition of the skeleton group recall that 
in an abelian gauge theory with gauge group $T$ the global electric symmetry is not restricted by any monopole background. For a non-abelian gauge theory with gauge group $G$ the global electric symmetry that can be realised in a monopole background is restricted but
always contains the maximal torus $T$ generated by the CSA of $G$. On the other, hand the magnetic charges can be identified with representations of the dual torus $T^*$. Hence the electric-magnetic symmetry in a gauge theory  with gauge group $G$ must contain $T\times T^*$. In the abelian case $T\times T^*$ is indeed the complete electric-magnetic symmetry group,  but in the non-abelian case we expect there to be a larger, non-abelian group containing $T\times T^*$.  In this section we will define such a group, and call it the skeleton group $S$. Our definition is such $T\times T^*$ is manifestly a subgroup of $S$, and $S$ equally manifestly a subgroup of  $G\times G^*$.   
 Furthermore,  the  irreducible representations of $S$ can be mapped to the magnetic, electric and dyonic charge sectors of non-abelian gauge theory.
                                                                       %
\subsection{Maximal torus and its dual}
\label{sect:maxtorus} 
The maximal torus $T$ is the maximal abelian subgroup of $G$ generated by $\mathfrak t$. 
In this section we are going to review an alternative definition which  can immediately be extended to give a  definition of $T^*$. This alternative description will  be the basis  for our 
discussion  of the Weyl group action on $T$ and $T^*$ in section \ref{sect:Weylaction}.

In section \ref{sect:lieconventions} we considered $\mathfrak t$ as a vector space over $\mathbb C$. However, if one declares the basis $\{H_\alpha\}$ of $\mathfrak t$ to be real, the real span of this basis defines a real vector space $\mathfrak t_{\mathbb R}$. Since any element  $t\in T$ can be written as $\exp(2\pi iH)$, with $H \in \mathfrak t_{\mathbb R}$, there is a surjective homomorphism
\begin{equation}
\label{eqn:homotT}
H \in \mathfrak t_{\mathbb R} \mapsto  \exp(2\pi iH) \in T.
\end{equation}
The kernel of this map is the set $\Lambda^*(G)$ and there is an isomorphism
\begin{equation}
T \sim \mathfrak t_{\mathbb R}/\Lambda^*(G).
\end{equation} 
As a nice consistency check of this isomorphism one can consider the irreducible representations and one will indeed find that  for $\mathfrak t_{\mathbb R}/\Lambda^*(G)$ these are labelled by elements of $\Lambda(G)$.

The dual torus $T^*$ is, by definition, a maximal abelian subgroup of $G^*$. Since  the coroots of $G^*$ can be identified with the roots of $G$, the real span $\mathfrak t^*_{\mathbb R}$ of the coroots of $G^*$  can be identified with  the real span   of the roots of $G$. By considering an analogous map to the one defined in  \eqref{eqn:homotT} we now find that $T^*$ is isomorphic $\mathfrak t^*_{\mathbb R}/\Lambda^*(G^*)$. In the case that $G^*$ is  the GNO dual of $G$ (so that $\Lambda^*(G^*)= \Lambda(G)$) we deduce that 
\begin{equation}
T^* \sim \mathfrak t^*_{\mathbb R}/\Lambda(G),
\end{equation}
which is consistent with the fact that the irreducible representations of the GNO dual group are labelled by elements of $\Lambda^*(G)$. 

A convenient way to  parametrise $T$ is as follows. Let $\widetilde G$ be the universal cover of $G$. The dual weight lattice $\Lambda^*(\widetilde G)$ for $\widetilde G$ equals the coroot lattice $\Lambda_{cr}$. A basis  of this lattice is the set of coroots $\{H_{\alpha_i}\}$ , where $\alpha_i$ are the simple roots of $G$. One thus finds that  the maximal torus $T_{\widetilde G}$  of ${\widetilde G}$ is explicitly parametrised by the set $\{H= \sum_{i=1}^r \theta_i H_{\alpha_i}\in \mathfrak t_{\mathbb R} ~|~ \theta_i \in [0,2\pi)\}$. Using the homomorphism \eqref{eqn:homotT},  we thus find that each element in $T_{\widetilde G}$ can uniquely be written as 
\begin{equation}
\label{eqn:paraT}
\exp\left(i\theta_iH_{\alpha_i}\right),
\end{equation}
with $\theta_i \in [0,2\pi)$.
If $G$ does not equal its universal covering group, equation \eqref{eqn:paraT} does not provide a unique parametrisation of $T$ in the sense that one still has to mod out the discrete group 
\begin{equation}
Z =\Lambda^*(G)/\Lambda_{cr}\subset T_{\widetilde G}.
\end{equation}
This follows from the fact that  $G =\widetilde G/Z$ and hence $T=T_{\widetilde G}/Z$.

Using analogous arguments we find that any element in $T^*$ can uniquely be represented as $H^*= \sum_{i=1}^r \theta^*_i H_{\alpha^*_i}$ up to an element in a discrete group $Z^*$. If $G^*$ equals the GNO dual of $G$, $Z^*$ is given by $\Lambda(G)/\Lambda_r$. 
\subsection{Weyl group action}
\label{sect:Weylaction}

The Weyl group is a subgroup of the automorphism group of the root system generated by the Weyl reflections
\begin{equation}
\label{weyldef}
w_\alpha: \beta \mapsto \beta - \beta(H_\alpha)\alpha.
\end{equation}
The action of the Weyl group can be extended  linearly to the whole root lattice, the weight lattice and $\mathfrak t^*$:
\begin{equation}
\label{invweyl1}
w_\alpha: \lambda \mapsto \lambda - \lambda(H_\alpha) \alpha,
\end{equation}
where we used the  notation \eqref{halphadef} and the natural duality between $\lambda \in t^*$ and $H_\alpha \in \mathfrak t$. Geometrically,  $w_\alpha$ is the reflection in the hyperplane in $\mathfrak t^*$   consisting of all $\lambda \in \mathfrak t^*$  which satisfy 
$\lambda(H_\alpha)=0$.

 The duality between $\mathfrak t$ and $\mathfrak t^*$ gives  rise to an  action of $w\in \mathcal W$ on $H\in\mathfrak t$, defined by the requirement that for all $\lambda\in \mathfrak  t^*$
 \begin{equation}
 \label{Weylpull}
 \lambda(w(H))= w^{-1}(\lambda)(H). 
 \end{equation}
 By definition, this action preserves the pairing between $\mathfrak t$ and $\mathfrak t^*$:
 \begin{equation}
 w(\lambda)(w(H))=\lambda(H).
 \end{equation}
 Explicitly, one finds for the elementary reflections (which satisfy $w_\alpha^2=1$, and therefore are their own inverses):
\begin{equation}
\label{invweyl2}
w_\alpha(H)= H- \alpha(H) H_\alpha.
\end{equation}
Note that  the fundamental definitions \eqref{invweyl1} and  \eqref{invweyl2} do not depend on the Killing form. The formula  
 \eqref{weyldef} can be expressed  in terms of the  inner product on the root space
\begin{equation}
w_\alpha: \beta \mapsto \beta - \frac{2\alpha\cdot\beta}{\alpha^2}\alpha.
\end{equation}
Similarly, equation\eqref{invweyl2}, specialised to 
  coroots, can be written as
\begin{equation}
w_\alpha(\beta^*)= \beta^*- \frac{2\beta^*\cdot \alpha^*}{(\alpha^*)^2}\alpha^*.
\end{equation}
Written in this way, the map $w_\alpha$ can be viewed as a reflection in the plane orthogonal 
to  the root vector $\alpha$.

The action of the Weyl group on $\mathfrak t$ induces an action on $T$ by exponentiation:
\begin{equation}
\label{eqn:WonT}
w\in \mathcal W: \exp\left( iH \right)\in T \mapsto \exp\left(i w(H) \right) \in T.
\end{equation}
Analogously one can define the action of the Weyl group on the dual torus:
\begin{equation}
\label{eqn:WonTdual}
w\in \mathcal W: \exp\left( i  H^*  \right) \in T^* \mapsto \exp \left(i w(H^*) \right) \in T^*.
\end{equation}

\subsection{Definition of the skeleton group}
\label{subsect:improvedskel}
The definition of the skeleton group in this section is motivated by the desire to recover the labelling of charge sectors via elements 
in  $(\Lambda\times\Lambda^*)/\mathcal W$  from the  
representation theory  of a subgroup $S$ of the  maximal (but non realisable) electric-magnetic symmetry $G\times G^*$.  It follows directly from the  representation theory of semi-direct products (which we will review in section \ref{subsect:reps}) that the group
 $\mathcal{W}\ltimes(T\times T^*)$, with the action of $\mathcal W$ on $T$ and $T^*$ defined as in \eqref{eqn:WonT} and \eqref{eqn:WonTdual}, does include representations with the required labelling. On the other hand, the Weyl group is, in general, not a subgroup of either $G$ or $G^*$,
 and therefore this semi-direct product does not fulfill our subgroup requirement.  Instead we define the skeleton group by the following, rather canonical construction. 
 
We use the notation $N_G(T)$ for the normaliser group of the maximal torus $T$ of $G$,  and $C_G(T)$ for the centraliser subgroup of $T$. By the definition of a CSA for compact Lie groups, we have $C_G(T)=T$.  
 Now, by one of the key theorems of Lie algebra theory, the quotient of the normaliser by the centraliser is isomorphic to the  Weyl group, so 
\begin{equation}
\frac{{ N}_G(T)}{{ C}_G(T)}\simeq {\mathcal W}.
\end{equation}
Since the Weyl group of $G$ and $G^*$ are the same, we also have
\begin{equation}
  \frac{N_{G^*}(T^*)}{ C_G^*(T^*)}\simeq {\mathcal W}.
\end{equation}
Hence there are canonical projections (surjective homomorphisms)
\begin{equation}
\label{lrproj}
\pi_{el}:  { N}_G(T)\rightarrow \mathcal W \leftarrow N_{G^*}(T^*) : \pi_{mag}  ,
\end{equation}
which map elements of ${ N}_G(T)$ and $N_{G^*}(T^*)$ to their associated Weyl elements.
We  now define the skeleton group in  terms of these   projections :
\begin{equation}
\label{newsdef}
S=\{ (y,y^*)\in  { N}_G(T) \times N_{G^*}(T^*) | \pi_{el}(y)=\pi_{mag}(y^*)\}.
\end{equation}
It may not be immediately obvious that the skeleton group is a group, but this follows from the fact that the projections $\pi_{el}$ and $\pi_{mag}$ are homomorphisms.  The requirement that $S$ be a subgroup of $G\times G^*$ is fulfilled since $S$ is a subgroup of   ${ N}_G(T) \times N_{G^*}(T^*) $, which in turn is clearly a subgroup of $G\times G^*$. It is manifest from the definition that, as a manifold, the skeleton group is a  fibre bundle over the Weyl group, with typical fibre $T\times T^*$.

Next we need to establish a  relation between the group structure of  $S$ and the semi-direct product  $\mathcal{W}\ltimes(T\times T^*)$. We are going to  do this by giving an explicit construction of $S$ in terms of generators, which will be useful in its own right.
In order to find generators, we are first going to consider purely electric and magnetic versions
of the skeleton groups, which are simply the normaliser groups of the maximal tori of $G$ and 
$G^*$: 
 \begin{equation}
 \label{selsmag}
S_{el}={ N}_G(T), \qquad S_{mag}={ N}_{G^*}(T^*).
\end{equation} 
Note that neither of  these groups is a subgroup of $S$, but that, by construction,  $S$ is a subgroup of $S_{el}\times S_{mag}$. 

Focusing on the electric version of the construction for definiteness, consider the projection  $\pi_{el}$ onto the Weyl group \ref{lrproj}. This projection can be realised explicitly by the conjugation action of $S_{el}$ on $T$: if $\pi_{el}(y)=w$ then, for any $t\in T$
\begin{equation}
yty^{-1}=w(t),
\end{equation}
 where $w(t)$ is the Weyl action of $w$ on elements of $T$ as defined in \ref{eqn:WonT}.  In order to understand the group structure of $S_{el}$ we  construct an explicit and discrete lift of the Weyl group  into $S_{el}$. The construction is a standard one, and is based  on lifts of generators of the Weyl group, which can be found, for example in  \cite{Fultonharris,Bouwknegt89}. As explained there,  the elements 
 \begin{equation}
 \label{eqn:defxalpha}
\u_\alpha = \exp\left(\frac{i\pi}{2}(E_\alpha + E_{-\alpha}) \right)
\end{equation} 
are lifts of the generators  $w_\alpha$ \eqref{invweyl2} of the Weyl group  in the sense that 
\begin{equation}
\label{projectdef}
\pi_{el}(\u_\alpha)=w_\alpha.
\end{equation}
The exponential map in \eqref{eqn:defxalpha} is the exponential map in $G$,  but the elements $\u_\alpha$ have to lie in $S_{el}=N(T)$ since they map $T$ (and $\mathfrak t$) into itself.
Thus one can define a lift $W_{el}$ of $\mathcal W$ into $S_{el}\subset G$ as the group generated by the elements $\u_\alpha$. Any element $\u\in S_{el}$ which also satisfies $\pi_{el}(\u)=w_\alpha$,
must satisfy
\[
\u  t \u^{-1}=\u_\alpha t \u_\alpha^{-1}\qquad \forall t\in T.
\]
It follows from the maximality of the CSA $\mathfrak t$ that $\u=t \u_\alpha t'$, for two elements $t,t'\in T$.  Since the Weyl reflections $w_\alpha$ generate the Weyl group we deduce from the surjectivity of the map $\pi_{el}$  that the elements $\u_\alpha$ together with all elements of $T$ generate $S_{el}$. Equivalently, we can say that the lift $W_{el}$ and the torus $T$ generate 
the electric skeleton group $S_{el}$. 

The lift $W_{el}$ of the Weyl group  generally contains elements that also lie in $T$, as noted above. The group
\begin{equation}
D_{el}= W_{el}\cap T
\end{equation} 
is an abelian group  because it is contained in $T$. It is also  a normal subgroup of  $W_{el}$
because it is the kernel of the  homomorphism
\begin{equation}
\label{welproj}
\pi_{el}:W_{el}\rightarrow \mathcal W,
\end{equation}
where we used  the notation $\pi_{el}$ also for the restriction to $W_{el}$ of the  map defined in  
\eqref{lrproj}.

The description of $S_{el}$ in terms of generators  will be  convenient for explicit constructions of this group in examples. In order to construct  representations, on the other hand, it is useful to note that the torus $T$ is a normal subgroup of $S_{el}$ and that $S_{el}$ has the structure of semi-direct product divided by the  abelian normal subgroup $D_{el}$:
\begin{equation}
\label{semiiso}
S_{el}\simeq \frac{W_{el}\ltimes T}{D_{el}},
\end{equation}
where the group multiplication in the semi-direct product is 
\begin{equation}
(u_1,t_1)(u_2,t_2)= (u_1u_2, t_1 u_1 t_2 u_1^{-1})
\end{equation}
 and   the action of $d\in  D_{el}$ is by diagonal  left-multiplication,  i.e.,   the quotient identifies
\[
(\u,t)\simeq (d \u ,d t)\in W_{el}\ltimes T.
\]
To show the ismorphism \eqref{semiiso} we  define $\phi:W_{el} \ltimes T \to G$ by
\begin{equation}
\label{phidef}
\phi(u,t) = t^{-1}u.
\end{equation} 
It is easy to check that $\phi$ is a homomorphism into $S_{el}=N_G(T)\subset G$, the normaliser of $T$.  The kernel of $\phi$ is precisely the set of elements $(d,d)\in W_{el}\ltimes T$, with necessarily $d\in D_{el}$. As a result, $S_{el}$ is  isomorphic to the image of $\phi$.
This image includes  the group generated by $\u \in W_{el}$ and $t\in T$ and hence it is all of $S_{el}$.

By a totally analogous construction one can define the lift $W_{mag}$ of the Weyl group into the magnetic group $G^*$,  and establish that $S_{mag}=N_{G^*}(T^*)$ is the group generated by
$W_{mag}$ and $T^*$ and that, with $D_{mag}=W_{mag}\cap T^*$,
\begin{equation}
S_{mag}\simeq \frac{W_{mag}\ltimes T^*}{D_{mag}}.
\end{equation}

Combining the descriptions of $S_{el}$ and $S_{mag}$ we can now derive two analogous descriptions of the skeleton group: one in terms of generators, which we can use for explicit descriptions, and a second as a quotient of a semi-direct product, which is useful for representation theory. For the first description, we combine the two lifts of the Weyl group into
\begin{equation}
\label{fibrationdef}
W=\{(\u,\u^*)\in W_{el}\times W_{mag} | \pi_{el}(\u)=\pi_{mag}(\u^*)\}.
\end{equation} 
This is clearly a discrete subgroup of $S$, while $T\times T^*$ is equally clearly an abelian subgroup of  $S$ (the fibre over the identity in the Weyl group). Now since any element in  
$S_{el}$  can be written as a product of   elements in $W_{el}$ and $T$, and similarly elements  $S_{mag}$ can be written as products of elements in $W_{mag}$ and $T^*$, it follows  that  $T\times T^*$  and elements in 
$W_{el}\times W_{mag}$  generate $S$, provided the latter  satisfy $\pi_{el}(\u)=\pi_{mag}(\u)$. Since this is precisely the defining condition of $W$ we deduce that $W$ and $T\times T^*$ generate $S$.

For the  semi-direct product description of $S$ we 
consider 
\begin{equation}
W\ltimes (T\times T^*),
\end{equation}
and introduce the abbreviation  $x=(\u,\u^*)$ for elements in $W$ as well as 
\begin{equation}
\label{pidef}
\pi(x)=\pi_{el}(\u)=\pi_{mag}(\u^*)
\end{equation}
for the Weyl element associated  to   $x=(\u,\u^*)$. Then $x$ 
 acts on $T\times T^*$ by  the  Weyl action of $\pi(x)$, which is explicitly given  by conjugation
\begin{equation}
\label{weyldouble}
\pi(x)(t,t^*)=(\u t\u^{-1}, \u^*t^*(\u^*)^{-1}).
\end{equation}

In  the semi-direct product some group elements occur in  both   $W$ and $T\times T^*$. To get rid of this redundancy  we define
\begin{equation} 
D = W\cap (T\times T^*),
\end{equation}
which is  an abelian,  normal subgroup of $W\ltimes (T\times T^*)$. Note that, in terms of the 
definition \eqref{fibrationdef} of $W$, the group $D$ is the fibre over the identity,  i.e.,  
\begin{equation}
D=\{(\u,\u^*)\in W_{el}\times W_{mag} | \pi_{el}(\u)=\pi_{mag}(\u^*)=1\}.
\end{equation} 
Since, by the discussion after \eqref{welproj} we have 
\[
D_{el}=\{\u\in W_{el} | \pi_{el}(\u)=1\}\quad \text{and}\quad  D_{mag}=\{\u^*\in W_{mag} | \pi_{mag}(\u^*)=1\},
\]
it follows that 
\begin{equation}
\label{dfactor}
D=D_{el}\times D_{mag}.
\end{equation}
Then we claim that, in analogy with the purely electric construction above, 
\begin{equation}
\label{olddef}
S \simeq  \frac{W\ltimes (T \times  T^*)}{D},
\end{equation}
with the quotient by t  diagonal  left-multiplication:
\[
(\u,\u^*,t,t^*)\simeq (d,d^*,d,d^*)(\u,\u^*,t,t^*), \qquad (d,d^*)\in D.
\]
To establish the isomorphism we define the homomorphism
\begin{equation}
\phi: W \ltimes T\times T^*\rightarrow  G\times G^*,\quad 
 (\u,\u^*,t,t^*) \mapsto ( t^{-1} \u,(t^*)^{-1} \u^*),
 \end{equation}
 which has, as kernel, the diagonal embedding of $D$ by which we divide when defining the 
 skeleton group. The image is easily seen to  lie inside $S\subset S_{el}\times S_{mag}$. Since it contains all generators of $S$ it is equal to $S$,  so that \eqref{olddef} follows by the first isomorphism theorem.

\subsection{The skeleton group for $SU(2)$}
\label{su2exsec}

Let us  illustrate the  definitions of the previous subsection for the case $G=SU(2)$. In this case the centre is $\mathbb Z_2$   and the dual group is
$G^*=SU(2)/\mathbb Z_2\simeq SO(3)$.  Our strategy for determining the skeleton group is to first determine the  lifts $W_{el}$ and $W_{mag}$ of the Weyl group, and to construct the electric, magnetic and full skeleton group from them.

Any  CSA  of $SU(2)$ is  one-dimensional and the Weyl group is generated by a single reflection  and  therefore  isomorphic to $\mathbb Z_2$.  We pick the diagonal matrices as the CSA of $SU(2)$,  and, since there is only one simple root, we  obtain the Cartan-Weyl basis for $SU(2)$ in the form 
\begin{equation}
H_\alpha=\begin{pmatrix}
1  &  \phantom{-}0 \\
0 &   -1  
\end{pmatrix}, \quad
E_\alpha= \begin{pmatrix}
0 &  \phantom{-}1 \\
0 & \phantom{-} 0
\end{pmatrix},
 \quad
 E_{-\alpha}=\begin{pmatrix}
0 &  \phantom{-}0 \\
1 &  \phantom{-}0
\end{pmatrix}.
\end{equation}
In terms of  Pauli matrices $\sigma_i$, $i=1,2, 3$,  we  find that
 the maximal torus  $T$ of $SU(2)$ is thus  $U(1)$-subgroup consisting of  diagonal elements
\begin{equation}
\label{u1elements}
\exp(i\varphi\sigma_3) = \begin{pmatrix}
e^{i \varphi} &  0 \\
0 &  e^{-i \varphi}
\end{pmatrix},
\end{equation}
with $\varphi\in [0, 2 \pi]$, 
and that  the element $\u$ of \eqref{eqn:defxalpha} is 
\begin{equation}
\label{xdef}
\u = \exp\left(i\frac{\pi}{2}\sigma_1\right) =i\sigma_1= \begin{pmatrix}
0 & \phantom{-} i \\
i & \phantom{-}  0
\end{pmatrix}.
\end{equation}
Clearly $\u^2=-1$ and $\u^4=1$, where $1$ stand for the identity matrix, so that $W_{el}\simeq \mathbb Z_4$. The group $D_{el}=W_{el}\cap T$ is precisely the centre of $SU(2)$ so $D_{el}\simeq \mathbb Z_2$.

Consider now  the electric skeleton group. It has the structure
\begin{equation}
S_{el} = \frac {\mathbb Z_4 \ltimes U(1)}{\mathbb Z_2},
\end{equation}
and can be realised explicitly  as a subgroup of $SU(2)$ as the group
generated  by the elements \eqref{u1elements} and the element $\u$ \eqref{xdef}.

Turning to the  magnetic skeleton group we repeat the above steps, but now working with the group  $SO(3)$ of rotations in 3-space. The maximal torus  $T^*$ is the group of rotations about the $3$-axis, while the   element 
 $\u^*$ that implements the Weyl reflection turns out to be the rotation by $\pi$ about the $1$-axis:
\begin{equation}
\label{wgen}
\u^*= \begin{pmatrix} 1 &\phantom{-} 0 &\phantom{-} 0 \\
0& -1&\phantom{-} 0 \\
0 & \phantom{-}0 & -1 \end{pmatrix}.
\end{equation}
Thus $W_{mag} \simeq  {\mathbb Z}_2$, and $D_{mag}$ is trivial so that 
$W_{mag}\simeq \mathcal W$ in this case.  The magnetic skeleton group is the group generated by arbitrary rotations about the 3-axis  and the $\pi$-rotation about the 1-axis. This group has the well-known semi-direct product structure
\begin{equation}
S_{mag} =  {\mathbb Z_2}\ltimes U(1)\simeq O(2).
\end{equation}

The geometrical picture of the magnetic skeleton group also sheds light on the electric skeleton group in this case, since  here $S_{el}$ and  $S_{mag}$ are related by
the standard  projection $ SU(2)\rightarrow SO(3)$: the element $\u$ is mapped to $\u^*$ under this projection, and $T$ is mapped to $T^*$. The electric skeleton group is  the pre-image of the magnetic skeleton group under this projection,  and thus a double cover of $O(2)$.

Finally we turn to the full skeleton group in this case. The lift $W$ of the Weyl group consists
of pairs of elements in $W_{el}\times W_{mag}$ that project to the same Weyl element. However, since $W_{mag}\simeq \mathcal W$ in this case, the magnetic element in the lift is uniquely determined once the electric element is picked. Thus $W$ is the group consisting of the four elements 
$W=\{(1,1),(\u,\u^*),(-1,1),(-\u,\u^*)\}\simeq \mathbb Z_4$, and the intersection $D$ with $T\times T^*$  is the two-element group $D=\{(1,1),(-1,1)\}$. 
The Skeleton group for $SU(2)$ therefore has the structure
\begin{equation}
\label{su2skel}
S= \frac {\mathbb Z_4 \ltimes (U(1)\times U(1))}{\mathbb Z_2}.
\end{equation}

The example has a number of features which are special.  For example, the
Weyl group turned out to be subgroup of the magnetic group $G^*$ in this case, 
and the magnetic skeleton group is simply a $\mathbb Z_2$-quotient of the electric skeleton group. The lift $W$ turned out be isomorphic to the electric lift $W_{el}$ of the Weyl group,
whereas in general it would be  bigger  group than  either $W_{el}$ or $W_{mag}$.
While the special features will not extend to general higher-rank groups, the example nonetheless illustrates the explicitness of our construction. Finally note that, if we had started with the group $G=SO(3)$ instead of $SU(2)$, electric and magnetic skeleton groups would be interchanged
  but the skeleton group would be isomorphic to \eqref{su2skel}.

\section{Representation theory}
\label{sect:repsskel}
In this section we discuss the representation theory  of the skeleton group and explain how the associated representation ring can be used to study  fusion rules of charge sectors in non-abelian gauge theory. The key result of  subsection \ref{subsect:reps} is the proof that there exists an irreducible representation of the skeleton group for every charge sector of non-abelian gauge theory as classified in section \ref{sect:chargesectors}.  The relationship between fusion rules for charge sectors and the  skeleton representation theory is discussed in general terms in  section \ref{subsect:fusionrules} and in much greater detail  for $G=SU(2)$  in section \ref{section:su2skel}. General results for   $G=SU(n)$  are gathered in appendix \ref{sect:impskelSU}. 

\subsection{Representations of the skeleton group}
\label{subsect:reps}
The  skeleton group is a subgroup of $G\times G^*$. This implies that representations of $G\times G^*$ decompose into irreducible representations of the skeleton group. In particular, viewing representations of $G$ or $G^*$ as representations of $G\times G^*$ where one factor is trivially represented, we can decompose purely electric or purely magnetic  representations into  irreducible representations of the skeleton group.
The skeleton group is thus an extension of $T\times T^*$ whose representation theory respects key features of the dyonic charge sectors. In this section we describe these  aspects of its representation theory in general terms and clarify the relation with representations of $G\times G^*$.

The representations of $S$ correspond precisely to the representations of $W\ltimes(T\times T^*)$ whose kernel contain the normal subgroup $D$. Since $W\ltimes(T\times T^*)$ is a semi-direct product its irreducible representations are labelled by an orbit and a centraliser representation \cite{Mackey49}. Here the relevant  orbits are  orbits of the $W$ action on  the character group of $T\times T^*$ , which is precisely given by $\Lambda(G)\times \Lambda(G^*)$. 
 Explicitly, $(\lambda,g)\in \Lambda(G)\times \Lambda(G^*)\subset \mathfrak t^*\times \mathfrak t$ is interpreted as a character, i.e.,  as a $\mathbb C$-valued function on $T\times T^*$  by writing $(t,t^*)\in T\times T^*$ as 
\begin{equation}
(t,t^*)=(\exp(iH),\exp(iH^*)), \qquad H\in \mathfrak t, H^*\in \mathfrak t^*,
\end{equation}   and defining
\begin{equation}
(\lambda, g):(t,t^*)\mapsto \exp(i\lambda(H) + i H^*(g)).
\end{equation}
In the following we will not differentiate notationally between the element $(\lambda, g)\in 
\Lambda(G)\times \Lambda(G^*)$ and the character defined by it. 

The action of $W$ on $T\times T^*$ is, by \eqref{weyldouble}, the diagonal Weyl group action 
on $T$ and $T^*$, as defined in equations \eqref{eqn:WonT} and \eqref{eqn:WonTdual}. The action of a Weyl element $w$ on $T\times T^*$ in turn induces  
 an action on  the character group via pull-back of the arguments with the inverse:
\begin{equation}
\label{pulll}
w(\lambda, g): (t,t^*)\mapsto (\lambda,g)(w^{-1}(t),w^{-1}(t^*))
 \end{equation}
or, comparing with \eqref{Weylpull}
\begin{equation}
\label{pullll}
w(\lambda, g):(t,t^*)\mapsto \exp(i w(\lambda)(H) + i w(H^*)(g)).
\end{equation}
Thus the action of $w$ on $(\lambda,g)$ interpreted as a character is the same as the action of 
$w$ on  $(\lambda,g)$  as an element of $\Lambda(G)\times \Lambda(G^*)$. Here and in the following we use simple juxtaposition to denote this action, as in equations \eqref{pulll}
and \eqref{pullll} above. 
The upshot of this discussion  is that  an irreducible representation of the skeleton group carries a label that corresponds to an $\mathcal W$ orbit $[\lambda,g]$ in 
$\Lambda(G)\times \Lambda(G^*)$. These labels are precisely the dyonic charge sectors of Kapustin \cite{Kapustin:2005py} as discussed in section \ref{subsect:dyonicsectors}. 

In order to give an explicit definition of the irreducible representations of the skeleton group let $[\lambda,g]$ denote the $\mathcal W$ orbit containing $(\lambda,g)$ and let $\gamma$ denote an irreducible representation of the centraliser $C_{(\lambda,g)} \subset W$ of $(\lambda,g)$. Now for any $(\mu,h)\in[\lambda,g]$, choose some $x_{(\mu,h)} \in W$ such that, in the notation of \eqref{pidef} for the Weyl element associated to $x$,  $\pi(x_{(\mu,h)})(\lambda,g)=(\mu,h)$.
We  define $V^{[\lambda,g]}_\gamma$ to be the vector space spanned by $\left\{\left|\mu ,h, e^\gamma_i\right>\right\}$, where $\left\{e^\gamma_i\right\}$ is a basis for the vector space $V_\gamma$ on which $\gamma$ acts. Using the standard  representation theory  of a semi-direct product \cite{Barut86} we find that the irreducible representation $\Pi_\gamma^{[\lambda,g]}$ of $W\ltimes(T\times T^*)$ acts on $V_\gamma^{[\lambda,g]}$ as follows:
\begin{eqnarray}
\label{semirep}
\Pi^{[\lambda,g]}_{\gamma}(x,t,t^*)\ket{\mu,h,\,v}=(\pi(x)(\mu,h))(t,t^*)\ket{\pi(x)(\mu, h), \gamma(x^{-1}_{\pi(x)((\mu,h))}\, x\, x_{(\mu, h)})v}.
\end{eqnarray}
These representations have the attractive property that the irreducible representations of $W\ltimes (T\times T^*)$ with trivial centraliser labels are in one-to-one relation with the electric-magnetic charge sectors. However,  not all of these representations are necessarily representations of $S$.

Representations of $W\ltimes (T\times T^*)$  are representations of $S$ if  the diagonal action of the  normal subgroup $D=W\cap (T\times T^*)$ is trivial, i.e.,  if 
\begin{eqnarray}
\label{precondid}
\Pi^{[\lambda,g]}_{\gamma}(d,d)\ket{\mu,h,\,v}=\ket{\mu,h,\,v}
\end{eqnarray}
for $d\in D$. 
It follows from the normality of  $D\subset W$ that  $x_{(\mu,h)}^{-1}dx_{(\mu,h)}= d'\in D$
and from $d\in T\times T^*$ that  $\pi(d)=1$. 
 Thus, the condition \eqref{precondid} becomes 
\begin{equation}
\label{condid}
(\mu,h)(d)\ket{\mu,h ,\gamma(d'}= \ket{\mu,h,\,v} \qquad \forall~\ket{\mu,h,\,v}\in V_\gamma^{[\lambda,g]}\quad\text{and}\quad \forall d\in D.
\end{equation}
Now we note that $(\mu,h)(d)=  (\lambda,g)(x_{(\mu,h)}^{-1}d x_{(\mu,h})
=(\lambda,g)(d')$. Hence  the condition \eqref{condid} becomes 
\begin{equation}
(\lambda,g)(d')\ket{\mu, h,\gamma(d')v} = \ket{\mu ,h, v} \qquad \forall~\ket{\mu,h,\,v}\in V_\gamma^{[\lambda,g]}\quad\text{and}\quad \forall d\in D.
\end{equation}
 As $d$ varies over $D$, the element $d'$ also sweeps out all of $D$.  Therefore the above condition is actually independent of $(\mu, h)$ and can be written as 
\begin{equation}
\label{eqn:gammaconstraint}
(\lambda,g)(d)\ket{\lambda, g, \gamma(d)v}= \ket{\lambda, g, v}\qquad \forall~\ket{v}\in V_\gamma, \quad d\in D,
\end{equation}
which is thus our condition for $D$ to  act trivially on $V_\gamma^{[\lambda,g]}$.

The condition \eqref{eqn:gammaconstraint} will be useful in example calculations, 
but one can  show that for every orbit $[\lambda,g]$ there exists at least one representation $\gamma$ 
which satisfies it  by   the following general argument.  The argument  exploits the fact that  
$S$ is a subgroup of $G\times G^*$, and that therefore
 all representations of  $G\times G^*$
 can be decomposed into irreducible representations of $S$. 
 This decomposition actually  furnishes the 
 decomposition of $G\times G^*$-representations into Weyl orbits, as can be seen by noting that states in $G\times G^*$-representations are labelled by  pairs $(\lambda, g)$ of electric and magnetic weights, and that an element $x \in W$ 
 acts on the labels via the Weyl element $\pi(x)$. Thus, in order to obtain an irreducible representation of $S$ which is labelled by the orbit $[\lambda,g]$ we can take any representation
 of $G\times G^*$ which contains a state with  weight $(\lambda,g)$, and decompose it into
 irreducible representations of $S$. Depending on the multiplicity of   the orbit $[\lambda, g]$
 in the Weyl orbit decomposition of the initial representation of $G\times G^*$,   we may obtain several irreducible representation of $S$ in the decomposition, and some of these may have 
 centraliser representations of dimension greater than one. However, we are guaranteed to obtain
 at least one irreducible representation of $S$ labelled by  $[\lambda, g]$, which is what we set out to show.

\subsection{Fusion rules}
\label{subsect:fusionrules}
We are now ready to address one of the key objectives of this paper, namely to explain how
the  representation ring of the skeleton group can be used to compute fusion rules for charge sectors in non-abelian gauge theory. We have seen that for every charge sector labelled by a electric-magnetic Weyl orbit $[\lambda,g]$ there exists an irreducible representation of the skeleton group. However, in general there are several such representations, which differ in their associated centraliser representations, some of which will generically have dimension bigger than one.  In this subsection  we show how to consistently   discard the centraliser representations,  
but in the next subsection we illustrate how one gains additional insights by retaining them. 
 
The representation ring of the skeleton group contains  a rule for  combining Weyl orbits
$[\lambda,g]$  since they  label the irreducible representations.  Formally one can thus derive a rule for combining  charge sectors by picking associated irreducible representations of the skeleton group, tensoring them  and ``forgetting" the label for the centraliser representation in 
the decomposition of the tensor product into irreducible representations. However, an equivalent
and efficient way of  computing the fusion rule is  to 
use a group ring constructed from the electric-magnetic charge lattice, as we will now explain.

Below we define a homomorphism,  called ``$\text{Char}$'',  from the representation ring of the skeleton group to the Weyl invariant part $\mathbb Z[\Lambda\times \Lambda^*]^\mathcal W$ of the group ring $\mathbb Z[\Lambda\times \Lambda^*]$ where $\Lambda\times \Lambda^*$ is the weight lattice of $T\times T^*$. This group ring has an additive basis given by the elements $e_{(\lambda,g)}$ with $(\lambda,g)\in \Lambda \times \Lambda^*$. The multiplication of the group ring is defined by $e_{(\lambda_1,g_1)}e_{(\lambda_2,g_2)}= e_{(\lambda_1+\lambda_2,g_1+g_2)}$. Finally,  the action of the Weyl group on the weight lattice induces an action on the group ring given by 
\begin{equation}
w\in \mathcal W: e_{(\lambda,g)}\mapsto e_{(w(\lambda),w(g))}.
\end{equation}
A natural basis for the ring $\mathbb Z[\Lambda\times \Lambda^*]^\mathcal W$ is the set of elements of the form
\begin{equation}
e_{[\lambda,g]}:=\sum_{(\mu,h)\in [\lambda,g]}e_{(\mu,h)},
\end{equation}
where $[\lambda,g]$ is a Weyl orbit in the weight lattice.

The homomorphism $\text{Char}$ from the representation ring of the skeleton group to\\
$\mathbb Z[\Lambda\times \Lambda^*]^{\mathcal W}$ is defined   via 
\begin{equation}
\text{Char}: \Pi^{[\lambda,g]}_\gamma \mapsto \text{dim}(V_\gamma)e_{[\lambda,g]}.
\end{equation}
Note that if $\gamma$ is a trivial centraliser representation or some other one-dimensional representation then $\text{Char}$ maps to a basis element of the group algebra.

$\text{Char}$ respects the addition and multiplication in the representation ring since
\begin{eqnarray}
&\text{Char}&: \Pi^{[\lambda_1,g_1]}_{\gamma_1}\oplus \Pi^{[\lambda_2,g_2]}_{\gamma_2} \mapsto \text{dim}(V_{\gamma_1}) e_{[\lambda_1,g_1]}+ \text{dim}(V_{\gamma_2})e_{[\lambda_2,g_2]}\\
&\text{Char}&: \Pi^{[\lambda_1,g_1]}_{\gamma_1}\otimes \Pi^{[\lambda_2,g_2]}_{\gamma_2} \mapsto \text{dim}(V_{\gamma_1}) \text{dim}(V_{\gamma_2})e_{[\lambda_1,g_1]}e_{[\lambda_2,g_2]}.
\end{eqnarray}
We can use this to retrieve the fusion rules for the dyonic charge sectors since the expansion of skeleton group representations in irreducible representations corresponds to expanding products in the Weyl invariant group ring into basis elements:
\begin{equation}
\label{eqn:ringfusion}
e_{[\lambda_1,g_1]}e_{[\lambda_2,g_2]}= \sum_{[\lambda,g]} N_{\lambda_1, \lambda_2, g_1, g_2}^{\lambda,g}e_{[\lambda,g]}.
\end{equation}

If one restricts to the purely electric or purely magnetic sectors one might hope  to retrieve the fusion rules of, respectively,  the full electric group  $G$ or the full magnetic group $G^*$.  However, as  noticed by Kapustin in \cite{Kapustin:2006hi}, equation \eqref{eqn:ringfusion} does not correspond to the decomposition of tensor products of $G$ representations.  Here the representation theory of the skeleton group $S$,  which also involves the centraliser representations,  offers additional information which  allows one to recover a greater part of the representation theory of $G$ and $G^*$. We will  illustrate this claim in the next section  for  the case $G=SU(2)$.

\subsection{Fusion rules for the skeleton group of $SU(2)$}
\label{section:su2skel}
Here we compute the complete set of irreducible representations and their fusion rules for the skeleton group of $SU(2)\times SO(3)$.  This allows us to predict fusion rules for the various sectors of the theory. We compare the skeleton fusion rules in the purely  magnetic and purely  electric sectors with those predicted by the full  magnetic and electric groups, but also compute fusion rules between magnetic and electric sectors, which go beyond the representation theory of the magnetic and electric groups. 
Finally we compare our computations with  results obtained by Kapustin and Saulina \cite{Kapustin:2007wm} using operator product expansions (OPE's) for dyonic  operators in twisted  $\mathcal N =4$ supersymmetric Yang-Mills theory. 

Our results suggest that the  skeleton group is a subgroup of the (yet to be determined) full symmetry object that governs the spectrum and fusion rules of the theory, and more importantly, a subgroup that can be realised in all electric-magnetic charge sectors of the theory.
We conjecture that the skeleton group is  the largest group which can be realized in all charge sectors simultaneously. If this is true, the fusion rules obtained from the skeleton group are all the information about the true fusion that one can hope to obtain within the usual framework of fusion described by the representation theory of a group.


Recalling the discussion of the skeleton group for $SU(2)$  in section \ref{subsect:improvedskel}
and using the notation introduced there, we note that
 irreducible representations of $S$ for $SU(2)$ correspond to a subset of irreducible representations of $\mathbb Z_4\ltimes (T\times T^*)$ which represent $D\simeq \mathbb Z_2$ trivially. This leads to a constraint on the centraliser charges and the electric charge as given by equation \eqref{eqn:gammaconstraint}.

If both the electric charge and magnetic charge vanish the centraliser is the group $\mathbb Z_4$ generated by  the element $\u$ defined in \eqref{xdef}.
The allowed centraliser representations are the two irreducible representations that represent $\u^2$ as $+1$. One of these representations is the trivial representation. This leads to the trivial representation of the skeleton group which we denote by $(+,[0,0])$. The  only non-trivial centraliser representations  map  $\u^2$ to $-1$ and gives a 1-dimensional  irreducible representation of the skeleton group which we shall denote by $(-,[0,0])$.

If either the electric or the magnetic charge does not vanish the orbit under the $\mathbb Z_4$ action has two elements and the centraliser group is $\mathbb Z_2 \subset \mathbb Z_4$ generated by $\u^2$. The irreducible representation of $\mathbb Z_2$ that satisfies equation \eqref{eqn:gammaconstraint} is uniquely fixed by the electric charge $\lambda$ labelling the equivalence class $[\lambda, g]$.  It is the trivial representation if the electric charge is even and it is the non-trivial representation if the electric charge is odd. We can thus denote the resulting irreducible skeleton group representation by $[\lambda,g]$ with $\lambda$ or $g$ non-vanishing. Note that these representations are 2-dimensional.

The electric-magnetic charge sectors appearing in the decomposition of a tensor product of irreducible representations of the skeleton group  can be found from the fusion rules of $\mathbb Z[\Lambda \times \Lambda^*]$ as discussed in section \ref{subsect:fusionrules}.  This gives the following fusion rules:
\begin{equation}
e_{[0,0]}e_{[0,0]} = e_{[0, 0]}, 
\end{equation}
\begin{equation}
e_{[0,0]}e_{ [\lambda, g]} = e_{[\lambda,g]},
\end{equation}
\begin{equation}
e_{[\lambda_1,g_1]}e_{[\lambda_2,g_2]} = e_{[\lambda_1+\lambda_2, g_1 +g_2]} + e_{ [\lambda_1-\lambda_2, g_1 -g_2]}. 
\end{equation}

Next consider the  full fusion rules of the skeleton group,  which  also take into account the centraliser representations. For all charges except $[0,0]$ the centraliser representations are uniquely determined. If we restrict to $[0,0]$ charges we obviously obtain $\mathbb  Z_4$ fusion rules. With $s,s_1,s_2\in\{\pm 1\}$ this leads to:
\begin{gather}
(s_1,[0,0])\otimes(s_2 ,[0,0]) = (s_1s_2,[0, 0]) \\
(s,[0,0])\otimes [\lambda, g] = [\lambda,g] \\
\label{eqn:skeldecomp3}
\left[\lambda_1,g_1\right]\otimes [\lambda_2,g_2] = [\lambda_1+\lambda_2, g_1 +g_2] \oplus [\lambda_1-\lambda_2, g_1 -g_2].
\end{gather}
If in the last line the electric-magnetic charges are parallel so that $[0,0]$ appears at the right hand side we have to interpret this as a 2-dimensional reducible representation. Its decomposition into irreducible representations can be computed via  characters, or, in the simple case at hand, using direct arguments. For later reference we note the general rule for  computing  fusion rules from the 
orthogonality of characters for  groups of the  semi-direct product form 
$H\ltimes N$,  with $H$ a finite group and $N$ abelian as is the case for $W\ltimes(T\times T^*)$.  The irreducible representations are then again labelled by orbits $[\sigma], [\eta],...$ and centraliser representations $\alpha, \beta...$. With the abbreviation  $a = \Pi^{[\sigma]}_\alpha$, $b= \Pi^{[\eta]}_\beta$ and $c= \Pi^{[\rho]}_\gamma$  the  fusion rules can obtained from  the 
orthogonality of characters according to 
\begin{eqnarray}
\left<\chi_{c},\chi_{a\otimes b}\right>\!\!
&=&  \int_{H\times N}  \chi_c(h,n))\chi^*_a(h,n)\chi^*_b(h,n) dhdn \nonumber 
   \\
&=&\sum_{\mu\in [\rho]}\sum_{\nu\in[\sigma]}\sum_{\zeta\in[\eta]}\delta_{\mu,\nu\zeta}\int_{H\times N} \delta_{h(\mu),\mu}\delta_{h(\nu),\nu} \delta_{h(\zeta),\zeta} \times  \label{eqn:fusionsemid}\\
&&\chi_\gamma(h_{\mu}^{-1}hh_{\mu})\chi^*_\alpha(h_{\nu}^{-1}hh_{\nu})\chi^*_\beta(h_{\zeta}^{-1}hh_{\zeta})dhdn,
\nonumber
\end{eqnarray}
where we have written summation over elements in $H$ as an integral, and  used $\chi_\alpha$
etc. to denote characters of the centraliser representations. 
For  the skeleton group of $SU(2)$ one finds
\begin{eqnarray}
\label{eqn:fussu2skel}
\left[\lambda,g\right]\otimes [\lambda,g] = [2\lambda, 2g] \oplus (-,[0,0]) \oplus (+,[0,0]).
\end{eqnarray}  
We would like to understand what the fusion rules obtained here teach us  about a possible extended electric-magnetic symmetry. The representations of such a symmetry should be uniquely labelled by the dyonic charges and should not carry additional quantum numbers. Moreover, the representations with vanishing magnetic charge should correspond to representations of the electric group. From this perspective the skeleton group representations  $(\pm,[0,0])$  should be interpreted  as part of odd dimensional representations of $SU(2)$. In this way one can at least reconstruct some of the fusion rules of $SU(2)$ in the magnetically neutral sector. As an example we consider equation \eqref{eqn:fussu2skel} with $\lambda$ equal the fundamental weight of $SU(2)$ and $g=0$:
\begin{equation}
\label{eqn:fussu2skelel}
[1,0]\otimes [1,0] = [2, 0] \oplus (-,[0,0]) \oplus (+,[0,0]).
\end{equation}
Since the skeleton group is always a subgroup of $G\times G^*$, we have, in this case, $S \subset SU(2)\times SO(3)$. Denoting representations of $SU(2)$ and $SO(3)$ by  underlined dimensions, we can thus
decompose a representation $\underline{n}\otimes \underline{m}$  of $SU(2)\times SO(3)$ into irreducible representations of $S$.
In particular the trivial representation $\underline{1}\otimes \underline{1} $ of $SU(2)\times SO(3)$ can be identified with the trivial representation of $S$ while the three-dimensional representation $\underline{3}\otimes \underline{1}$ falls apart into $[2,0] \oplus (-,[0,0])$, with $2$ equal to the highest weight of the triplet representation.  Interpreted in this way,  
the equation \eqref{eqn:fussu2skelel} is tantamount to the  familiar rule for combining two doublets of the electric group $SU(2)$: 
\begin{equation}
\label{eqn:doubletsquared}
\underline{2}\otimes \underline{2} = \underline{3}\oplus \underline{1},
\end{equation}
with the magnetic group represented trivially.

One could try to push this line of thought further to include the case  $g\neq 0$. In particular,  it would be interesting  to see if the combination of two dyonic charge sectors with equal magnetic charges can lead to purely electric sectors which can be viewed as $SU(2)$ representations.  Since the right hand side of equation~\eqref{eqn:fussu2skel} does not contain the  electric sector  $[2\lambda,0]$  when $g\neq 0$
it looks like the skeleton group does not allow for  such a possibility.  However, this conclusion may be premature. The skeleton group is only expected to be a subgroup of the full symmetry object governing dyonic charges and it is likely that a number of skeleton group representations must be combined into a representation of the full symmetry object. For purely electric and magnetic charges, we know how to do this, namely using restriction from $G\times G^{*}$ to $S$ (or induction from $S$ to $G\times G^{*}$). For general dyonic sectors we do not know which skeleton group representations should be combined because we do not know the full symmetry. However, it is very likely that combinations are necessary and by tensoring combinations, such as for instance $\left[\lambda,g\right]\oplus \left[\lambda,-g\right]$ it is not difficult to obtain purely electric skeleton representations with nonzero electric weights on the right hand side of an equation analogous to~\eqref{eqn:fussu2skel}.

A different approach to finding a unified description of an electric group $G$ and a magnetic group $G^*$ is to consider the OPE algebra of mixed Wilson-'t Hooft operators. Such operators are labelled by the dyonic charge sectors as explained by Kapustin in \cite{Kapustin:2005py}. Moreover, the OPEs of Wilson operators are given by the fusion rules of $G$ while the OPEs for 't Hooft operators correspond to the fusion rules of $G^*$. These facts were used by Kapustin and Witten \cite{Kapustin:2006pk} to prove that magnetic monopoles transform as $G^*$ representations in  a topological version  of $\mathcal N=4$ supersymmetric Yang-Mills theory. 
It is thus natural to ask what controls the product of mixed Wilson-'t Hooft operators. The answer must somehow unify the representation theory of $G$ and $G^*$. Consequently one might also expect it to shed some light on the fusion rules of dyons.

For a twisted $\mathcal N=4$ SYM theory with gauge group $SO(3)$ products of Wilson-'t Hooft operators have been computed by Kapustin and Saulina \cite{Kapustin:2007wm}. In terms of dyonic charge sectors they found for example:
\begin{equation}
\label{eqn:ope1}
[n,0]\cdot[0,1]= \sum_{j=0}^{n} [n-2j,1].
\end{equation}
This rule can  be understood from the fusion rules of the skeleton group $S$ for $SO(3)\times SU(2)$ (which is the same as the skeleton group of $SU(2)\times SO(3)$, but with the electric and magnetic interpretation interchanged) : we interpret the sector $[n,0]$  as the $SO(3)\times SU(2)$ representation $\underline{n+1}\otimes \underline{1}$,   decompose into irreducible representations of the  skeleton group $S$  and then combine
it with the magnetic sector $[0,1]$ using the fusion rule of the full skeleton group.  The decomposition into a sum of irreducible representations of $S$ is 
\begin{equation}
\label{eqn:su2skeletonbranchingeven}
\underline{n+1}\otimes \underline{1}=\bigoplus_{j=0}^{\frac{n}{2}-1} [n-2j,0] +(s,[0,0]),
\end{equation}
with  the centraliser label $s$ in \eqref{eqn:su2skeletonbranchingeven} uniquely determined by $n$. The 't Hooft operator labelled by $[0,1]$ can be  identified with  the irreducible representation $[0,1]$ of the skeleton group. Similarly for the Wilson-'t Hooft operators appearing at the right hand side of equation \eqref{eqn:ope1} there is also a unique identification with skeleton group representations. Thus  we reproduce  the right hand side of equation \eqref{eqn:ope1} via  the decomposition of the tensor products of $[0,1]$ with the reducible representations \eqref{eqn:su2skeletonbranchingeven} into irreducible representation of the skeleton group:
\begin{eqnarray}
&&\left(\bigoplus_{j=0}^{\frac{n}{2}-1} [n-2j,0] +(s,[0,0])\right)\otimes [0,1]\\
&=& \bigoplus_{j=0}^{\frac{n}{2}-1} [n-2j,1]\oplus  \bigoplus_{j=0}^{\frac{n}{2}-1} [-n+2j,1]\  \oplus [0,1]\\
&=& \bigoplus_{j=0}^{n} [n-2j,1],
\end{eqnarray}
where we  made repeated use of \eqref{eqn:skeldecomp3}.

A second product rule obtained in \cite{Kapustin:2007wm}, which is consistent with the results of \cite{Kapustin:2006pk}, is a fusion rule for  purely magnetic charge sectors  in the theory with gauge group $SO(3)$:
\begin{equation}
\label{eqn:ope2}
[0,1]\cdot[0,1]= [0,2] + [0,0].
\end{equation}
This product is also easy  to understand from the fusion rules of the skeleton group $S$ of $SO(3)\times SU(2)$.  In terms of irreducible representations of $S$ we have
\begin{equation}
[0,1]\otimes[0,1] = [0,2]\oplus (-,[0,0]) \oplus (+,[0,0]). 
\end{equation}
As in the case of equation \eqref{eqn:fussu2skelel} we  argue  again that  the $S$-representations $(-,[0,0])$  and $[0,2]$ should be  combined and interpreted as making up magnetic sector $[0,2]$, 
i.e.,~the triplet representation of the magnetic group $SU(2)$.

Finally consider the following 
OPE product rule found in \cite{Kapustin:2007wm}:
\begin{equation}
\label{eqn:ope3}
[2n,1]\cdot[0,1]= [2n,2] + [2n,0] - [0,0] - [2n-2,0].
\end{equation}
Negative terms can occur naturally in the $K$-theory approach  used in \cite{Kapustin:2007wm}.
In our tensor product approach we only have positive terms, and  in the case at hand these  follow 
 from equation \eqref{eqn:skeldecomp3}:
  \begin{equation}
\label{eqn:skelfusion3}
[2n,1]\cdot[0,1] = [2n,2] \oplus [2n,0].
\end{equation}
One observes that the terms missing in this last equation correspond to the terms in equation \eqref{eqn:ope3} with a minus sign.
We conclude that fusion rules of the skeleton group are to some extent consistent with the OPE algebra discussed by Kapustin and Saulina. The advantage of their approach is first that there is never need to restrict the gauge groups to certain subgroups as we effectively do with the skeleton group. Nonetheless, because of the occurrence of negative terms, the OPE algebra cannot easily be interpreted as a set of physical fusion rules for dyons.
\section{S-duality}
\label{sect:sduality}
In this section we consider, for the first time in this paper, a specific class of gauge theories, namely 
$\mathcal N=4$ supersymmetric Yang-Mills theories.
We review the standard implementation of S-duality, 
and define an S-duality action on the skeleton group representations which commutes with the fusion rules of the previous section.
\subsection{S-duality for simple Lie groups}
In $\mathcal N=4$ SYM theory with unbroken gauge group $G$,  S-duality  acts on the complex coupling constant $\tau= \frac{\theta}{2\pi}+ \frac{4\pi i}{e^2}$ and the electric-magnetic charges.
The action of S-duality group on the electric-magnetic charges is discussed in the general case  in \cite{Girardello:1995gf,Dorey:1996jh};  see also \cite{Kapustin:2005py} for a  succinct summary.
This action makes use of the Killing metric on the $\mathfrak t$, so our brief review of it is the 
first occasion in this paper  where the Killing metric is used in an essential way.

 First we choose the short coroots to have length $\sqrt{2}$, i.e.,  $
\left< H_\alpha, H_\alpha \right> = 2.$
Adopting the conventions and notation of   \cite{Kapustin:2005py} we define  a  linear map $\ell$ acting on the CSA of $G$ and its dual 
\begin{equation}
\label{eqn:metricdual}
\begin{split}
&\ell: H_\alpha \in \mathfrak t \mapsto H_\alpha^\star = \frac{\left<H_\alpha, H_\alpha\right>}{2}\alpha \in \mathfrak t^*.\\
&\ell ^{-1}: \alpha \in \mathfrak t^* \mapsto \alpha^\star = \frac{2H_\alpha}{\left<H_\alpha, H_\alpha\right>} \in \mathfrak t.
\end{split}
\end{equation}
and use it to define the following actions
\begin{eqnarray}
C: \tau \mapsto \tau \quad  && (\lambda, g) \mapsto (-\lambda, -g) \\
T: \tau \mapsto  \tau + 1  \quad &&(\lambda, g) \mapsto (\lambda - g^\star, g) \\
S: \tau \mapsto - \frac{1}{\tau} %
                                     \quad  &&(\lambda,g) \mapsto (g^\star, -\lambda^\star).
\end{eqnarray}
One can check that $C^2=1$, $S^2=1$ and $(ST)^3=C$.  The elements $T$ and $S$ generate the group  $SL(2,\mathbb Z)$ and $C$ is the  non-trivial element of its centre. 
Unfortunately, the electric-magnetic charge lattice $\Lambda(G)\times \Lambda(G^*)$ is in general not mapped onto itself under the action of $SL(2,\mathbb Z)$. However, as explained in section \ref{sect:chargesectors}, it is natural in an $\mathcal N=4$ gauge theory with smooth monopoles to take both $G$ and $G^*$ to be adjoint groups and thereby restrict the electric charges to the root lattice and the magnetic charges to the coroot lattice.  Then the  lattice $\Lambda_r\times \Lambda_{cr}$ is invariant under some subgroup of $SL(2,\mathbb Z)$. To see this note that a long coroot $H_\alpha$  is mapped to a multiple of $\alpha$ since the length-squared of a long coroot is an integral  multiple of the length-squared for a short coroot. Consequently, the image of $\Lambda_{cr}$ under $\ell$ is contained in the root lattice $\Lambda_r$ of $G$. Next we need to check if $\ell^{-1}$ maps the root lattice  of $G$  into the coroot lattice.  This is clearly not the case if $ G$ has long and short roots, since
the length-squared of the image of a  long root  has length-squared smaller than $2$. Hence the root lattice is mapped into the coroot lattice by $\ell^{-1}$  only if $G$ is simply-laced.

 In the non-simply laced case the action of the generator $S$ does not leave $\Lambda_r\times\Lambda_{cr}$ invariant. However, as shown in \cite{Girardello:1995gf,Dorey:1996jh}
   one can still consider the transformation $ST^qS$ which acts as
\begin{equation}
ST^qS: (\lambda,g)\to(-\lambda,-q\lambda^\star-g).
\end{equation}
For $q$ sufficiently large $q\lambda^\star$ is always an element of the coroot lattice, hence there is a subgroup $\Gamma_0(q)\subset SL(2,\mathbb Z)$ that generated by $C,T$ and $ST^qS$ that leaves $\Lambda_r\times \Lambda_{cr}$ invariant. The largest possible duality group for e.g. $SO(2n+1), Sp(2n)$ and $F_4$ is $\Gamma_0(2)$ while for $G_2$ it is $\Gamma_0(3)$. 


\subsection{S-duality on charge sectors}
We have seen above that there is an action of $SL(2,\mathbb Z)$ (or at least an action of a subgroup $\Gamma_0(q)$) if we restrict the electric-magnetic charge lattice to $\Lambda_r \times \Lambda_{cr}$.  The restriction of the charge lattice also defines a restriction of the dyonic charges sectors to $(\Lambda_r \times \Lambda_{cr})/\mathcal W$.  One can see that S-duality has a well-defined action on the charge sectors  by noting  that the action of the generators of $SL(2,\mathbb Z)$ commute with the diagonal action of the Weyl group \cite{Kapustin:2005py}.  This is obvious for $C$ since $wC(\lambda, g)= w(-\lambda, -g)= (-w(\lambda), -w(g)) = Cw(\lambda, g)$. For $T$ and $w\in \mathcal W$ we have: $wT(\lambda, g)= w(\lambda + g^\star, g) = (w(\lambda) + w(g^\star), w(g)) = (w(\lambda) + w(g)^\star, w(g))= T(w(\lambda), w(g))= Tw(\lambda,g)$. Finally for $S$ we have $wS(\lambda,g) = w(-g^\star, \lambda^\star) = (-w(g)^\star, w(\lambda)^\star)= Sw(\lambda, g)$.
\subsection{S-duality and skeleton group representations}
We would like to show that the 
action of the duality group on the dyonic charge sectors can be extended  to the set of representations of the skeleton group. The latter carry labels  for  centraliser representations of the lifted Weyl group $W$ in addition of the dyonic charge sector labels. We shall  show that one obtains a well-defined action if one assumes that S-duality acts trivially on the centraliser representations, and that this action commutes with the fusion rules of the skeleton group.   Before we do this, note  that we are not considering all representations of the skeleton group but only those that correspond to the root and coroot lattice. Effectively, we have thus modded the skeleton group out by a discrete group.

To show that the S-duality action is well-defined  we first observe that the  action of $C,T$ and $S$, and hence also the action of the duality group commutes with the action of the lifted Weyl group.  This follows from the fact that the duality group commutes with the Weyl group,  as shown in the previous section.  

Next we show that  the centraliser subgroup  in $W$ is invariant under the action of the duality group on the electric and magnetic charge, 
using the notation from section \eqref{sect:repsskel}.
Since the action of $\mathcal W$ and thus also $W$ on the electric-magnetic charges is linear it follows that charge conjugation does not change the centraliser. The fact that $T$ leaves the centraliser group $C_{(\lambda,g)}\subset W$ invariant is seen a follows: let $C_g\subset W$ be the centraliser of $g$ so that for every $w\in C_g$ $w(g)=g$.  The centraliser of $(\lambda,g)$ consists of elements in $w\in C_g$ satisfying $w(\lambda)= \lambda$. Similarly the elements  $w \in C_{(\lambda+g^\star,g)}$ satisfy $w(g)=g$ and thus $w(g^\star)=g^\star$. Finally one should have $w(\lambda +g^\star)= \lambda +g^\star$. But since $w(\lambda+ g^\star)= w(\lambda) + w(g^\star)$ one finds that $w$ must leave $\lambda$ invariant. Hence $C_{(\lambda+ g^\star,g)}= C_\lambda \cap C_g= C_{(\lambda,g)}$. Similarly the action of $S$ is seen to leave the leave $C_{(\lambda,g)}$ invariant since $C_{\lambda^\star}= C_\lambda$ and $C_{-g^\star}= C_{g}$ so that $C_{-g^\star}\cap C_{\lambda^\star}=  C_{\lambda}\cap C_{g}$.

An irreducible representation of the skeleton group is defined by an orbit in the electric-magnetic charge lattice and an irreducible representation of the centraliser in $W$ of an element in the orbit. Since the $SL(2,\mathbb Z)$ action commutes with the action of the lifted Weyl group, a $W$ orbit is mapped to another $W$ orbit. We define the centraliser representation to be invariant under the duality transformation. This is consistent because the centraliser subgroup itself is invariant under $SL(2,\mathbb Z)$. We thus find that an irreducible representation of the skeleton group is mapped to another irreducible representation under the duality transformations.

Finally we prove that S-duality transformations respect the fusion rules of the skeleton group. The claim is that if for irreducible representations $\Pi_a$ of the skeleton group one has 
\begin{equation}
\Pi_a \otimes \Pi_b =  n_{ab}^c \Pi_c,
\end{equation}
then for any element $s$ in the duality group one should have
\begin{equation}
\label{eqn:sdualfusion}
\Pi_{s(a)} \otimes \Pi_{s(b)} =  n_{ab}^c \Pi_{s(c)}.
\end{equation}
We can prove this equality by inspection of  the general formula \eqref{eqn:fusionsemid}. First we note that since $s$ commutes with the lifted Weyl group we have, for any $(\mu',h')\in [s(\lambda,g)]$, that  $(\mu',h')=s(\mu,h)$ for a unique $(\mu,h)\in [\lambda,g]$. Thus the summation over the orbits $[\lambda,g]$ and $[s(\lambda,g)]$ is equivalent. Next we see that since $s$ is an invertible linear map on the dyonic charges $s(\mu_3,h_3)= s(\mu_1,h_1)+ s(\mu_2,h_2)$ if and only if $(\mu_3,h_3)= (\mu_1,h_1)+ (\mu_2,h_2)$. Similarly we find that for any 
$x\in W$,   $\pi(x) s(\mu,h)= s(\mu,h)$ if and only if $\pi(x)(\mu,h)=(\mu,h)$, where we again use the notation for the Weyl action on the charge lattice introduced before \eqref{semirep}. Finally we note that in terms of  $x_{(\mu,h)}\in W$ which satisfies $x_{(\mu,h)} (\lambda,g)= (\mu,h)$ we have  $\pi(x_{(\mu,h)})s(\lambda,g)= s(\pi(x_{(\mu,h)})(\lambda,g))= s(\mu,h)$ and hence $x_{s(\mu,h)}= x_{(\mu,h)}$. With our conjecture that the  S-duality action does not affect the centraliser charges we now  conclude 
directly from  \eqref{eqn:fusionsemid} that 
\begin{equation}
\left<\chi_{c},\chi_{a\otimes b}\right>= \left<\chi_{s(c)},\chi_{s(a)\otimes s(b)}\right>.
\end{equation}
This proves \eqref{eqn:sdualfusion}.
\section{Conclusion and outlook}
\label{outlook}

In this paper we proposed the skeleton group as a candidate for a non-abelian electric-magnetic symmetry in gauge theories with an unbroken non-abelian gauge  group. The definition of the skeleton group  only uses data naturally associated to the unbroken gauge group and its GNO dual. We demonstrated that the skeleton group allows one to study fusion rules of electric, magnetic and dyonic charge sectors and that it is compatible with S-duality in  $\mathcal N$=4 supersymmetric Yang-Mills theory. However, many aspects and potential applications of the skeleton group remain unexplored here.  In particular,  given the generically tight connection between  symmetry  and phase structure in field theory,   we expect  the skeleton group to play a key role in studying   phases and phase transitions in non-abelian gauge theories. In this final outlook section
 we briefly sketch how how such a study might proceed.

We begin by recalling  an interesting proposal of 't Hooft \cite{'tHooft:1981ht}. In order to get a handle on non-perturbative effects in gauge theories, like chiral symmetry breaking and confinement, 't Hooft introduced the notion of \textit{non-propagating gauges}. An important example of such a non-propagating gauge is the so-called \textit{abelian gauge}. In this gauge a non-abelian gauge theory can be interpreted as an abelian gauge theory (with the abelian gauge group equal to the maximal torus  of $G$) with monopoles in it. This has led to a host of interesting approximation schemes to tackle the aforementioned non-perturbative phenomena which remain elusive from a first principle point of view, see, e.g., \cite{Kronfeld:1987vd,Kronfeld:1987ri,Smit:1989vg,Shiba:1994ab}.

The skeleton group can be used to  generalise   't Hooft's proposal, from an abelian to a minimally non-abelian scheme. Instead of the maximal torus one uses the skeleton group   as a  residual symmetry in a gauge which one might call the  skeleton gauge. The attractive feature is that this generalisation does not affect the continuous part of the residual gauge  symmetry after fixing. It is still abelian, but the generalisation adds (non-abelian) discrete components to that residual symmetry. This implies that in the skeleton gauge the non-abelian features of the gauge theory manifest themselves through topological interactions only, and that makes them manageable. The effective theories we end up with are  generalisations of Alice electrodynamics \cite{Kiskis:1978ed,Schwarz:1982ec,Alford:1990ur}. In this sense the effective description of the non-abelian theory with gauge group $G$ in the skeleton gauge is a merger of an abelian gauge theory and a (non-abelian) discrete gauge theory \cite{Preskill:1990bm,Propitius}.

Working in  the skeleton gauge  we expect to be able to  answer kinematic questions concerning different phases and possible transitions between them. For this purpose it is of the utmost importance to work in a scheme where one can compute the fusion rules involving electric, magnetic and dyonic sectors. This is evident in the abelian case where the fusion rules are very simple: if there is a condensate corresponding to a particle with a certain electric or magnetic charge then any particle with a multiple of this charge can consistently be thought of as absorbed by the vacuum. For confinement we know that if two electric-magnetic charges do not confine then the sum of these charges will also not confine.
Given the fusion rules predicted by the skeleton group we can
therefore, at least in principle, use an approach analogous to that
employed in \cite{Bais:2002pb} to analyse all phases that emerge from
generalised Alice phases by condensation or confinement. We intend to
report on this analysis in a future publication.

\acknowledgments
This work is part of the research programme of the `Stichting voor Fundamenteel
Onderzoek der Materie (FOM)', which is financially supported by the `Nederlandse
Organisatie voor Wetenschappelijk Onderzoek (NWO)'.
\appendix
\section{Skeleton group for $SU(n)$}
\label{sect:impskelSU}
Below study  the skeleton group and its irreducible representations in some detail for $G=SU(n)$. Skeleton groups for the other classical Lie groups are discussed in \cite{Kampmeijerphd}  but the discussion there is based on a slightly different definition. 

We shall start by identifying the electric  lift $W_{el}$ of the Weyl group. For the maximal torus $T$ of $SU(n)$, we take the subgroup of diagonal matrices. The length of the roots is set to $\sqrt{2}$. 
The raising and lowering operators for the simple roots are the matrices given by $(E_{\alpha_i})_{lm} = \delta_{li}\delta_{m,i+1}$ and  $(E_{-\alpha_i})_{lm} = \delta_{l,i+1}\delta_{m,i}$. 
From this one finds that $\u_{\alpha_i}$ as defined in equation \eqref{eqn:defxalpha} is given by:
\begin{equation} 
(\u_{\alpha_i})_{lm}= \delta_{lm}(1-\delta_{li}
-\delta_{l,i+1})+ i(\delta_{li}\delta_{m,i+1}+ \delta_{l,i+1}\delta_{mi}).
\end{equation}
From now on we abbreviate $\u_{\alpha_i}$ to $\u_i$. One easily shows that \begin{equation}
\u_i^4=1, \qquad [\u_i,\u_j]=0~\text{for}~|i-j|>1, \qquad \u_i\u_{i+1}\u_i= \u_{i+1}\u_i\u_{i+1}.
\end{equation}
As it stands, this is not the complete set of relations for $W_{el}$. However, one may show that $W_{el}$ is fully determined if we add the relations
\begin{equation}
(\u_i\u_{i+1})^3=1.
\end{equation}
This also makes contact with the presentation of the normaliser of $T$ obtained by Tits \cite{Tits66a,Tits66b}.

We shall now determine the group $D_{el}$. Note that the elements $\u_i^2\in W_{el}$ are diagonal and of order 2. In fact, we have $(\u_i^2)_{lm}=\delta_{lm}(1- 2\delta_{li}-2\delta_{l,i+1})$. One thus sees that the group $K$ generated by the $\u_i^2$ is just the group of diagonal matrices with determinant 1 and diagonal entries equal to $\pm 1$. Since its elements are diagonal we have $K\subset T$ and hence $K\subset D_{el}=W_{el}\cap T$. As a matter of fact 
$K=D_{el}$. To prove this, we  recall  that, as explained before equation \eqref{welproj},
$D_{el}$ is the kernel of the projection of $W_{el}$ to the Weyl group, which, in the case at hand, is the permutation group $\mathcal S_n$. However, it is easy to see 
  that $W_{el}/K$  already satisfies the relations of the permutation group, which are the same as the relations for the $\u_i$ above, but with $\u_i^2=1$. Thus $K$ is precisely the group by which we need to divide $W_{el}$ to obtain $\mathcal W$,  hence  $D_{el}=K$. In view of the explicit generators given above it is easy to see  that $D_{el} \simeq \mathbb Z_2^{n-1}$.

Finally, the electric skeleton group is the group generated by  the maximal torus $T\simeq (U(1))^{n-1}$, consisting of  diagonal elements in $SU(n)$,  and the elements  $u_i\in W_{el}$. It follows from the above, that this group has the structure
\begin{equation}
S_{el}\simeq \frac{W_{el}\ltimes (U(1))^{n-1}}{\mathbb Z_2^{n-1}}.
\end{equation}

Next consider the magnetic  skeleton group. The magnetic group is $G^*=SU(n)/Z_n$, where $Z_n\simeq \mathbb Z_n$ is the centre of $SU(n)$, consisting of the identity matrix multiplied by an $n$th root of unity.  To construct $S_{mag}$ we thus only need to divide by this subgroup in the appropriate places. However, it is easy to see from the explicit expression for the generators $\u_i$ of $W_{el}$ they and their powers always have real numbers on the diagonal, so  that 
\begin{equation}
W_{el}\cap Z_n =\left\{
\begin{array}{l l} \{ 1 \} \qquad &\text{if $n$ is odd}\\
\{1,-1\}&\text{if $n$ is even.} 
\end{array}
\right. 
\end{equation}
Thus, the magnetic skeleton group is 
\begin{equation}
\label{magskelsun}
W_{mag} \simeq \left\{
\begin{array}{l l} W_{el} \qquad &\text{if $n$ is odd}\\
W_{el}/\mathbb Z_2&\text{if $n$ is even},
\end{array}
\right. 
\end{equation}
where  $\mathbb Z_2=\{1,-1\}$.
Explicitly, we can think of the generators $\u_i^*$ of the dual skeleton group as cosets 
$\u_i Z_n$ in $SU(n)/Z_n$. These cosets will contain $\pm \u_i$ in the case where $n$ is even, therefore identifying those elements.
It follows that $D_{mag}$ is isomorphic to $D_{el}$ if $n$ is odd, and isomorphic to the quotient
$D_{el}/\mathbb Z_2\simeq \mathbb (Z_2)^{n-2}$ if $n$ is even.  Since the magnetic torus $T^*=T/C_{n}$ is also isomorphic to $(U(1))^{n-1}$ we have the structure
\begin{equation}
S_{mag}\simeq \left\{
\begin{array}{l l} S_{el} \qquad &\text{if $n$ is odd}\\
((W_{el}/\mathbb Z_2) \ltimes (U(1))^{n-1})/\mathbb Z_2^{n-2}
&\text{if $n$ is even}.
\end{array}
\right. 
\end{equation}
It is interesting to note that electric and magnetic skeleton groups are isomorphic for odd $n$ even though the full electric and magnetic groups are not. 

The full skeleton group consists, by definition \eqref{newsdef} and the remark after \eqref{selsmag}  of pairs of elements $(y,y^*)\in S_{el}\times S_{mag}$ which project to the same Weyl element, i.e., the same permutation in $\mathcal S_n$ in the current class of examples. It can constructed explicitly for any given $n$, using the generators given up. However, we have not been able to give any characterisation of this group  for $SU(n)$ which goes beyond the formulations given for the general case in the main text.

In order to determine the representations of $S$ for $SU(n)$ we need to solve \eqref{eqn:gammaconstraint} and hence we need to describe how $D$ is represented on a state $\ket{\lambda}$ in an arbitrary representation of $SU(n)\times SU(n)/Z_n$. This turns out to be surprisingly easy. Recalling \eqref{dfactor} the factorisation $D=D_{el}\times D_{mag}$
we can treat the electric and magnetic side separately.

The generating element $\u_i^2$ of $D_{el}$ acts as the non-trivial central element of the $SU(2)$ subgroup in $SU(n)$ that corresponds to $\alpha_i$. Now let $(\lambda_1,\dots, \lambda_{n-1})$ be the Dynkin labels of the weight $\lambda$. Note that $\lambda_i$ is also the weight of $\lambda$ with respect to the $SU(2)$ subgroup corresponding to $\alpha_i$. Recall that the central element of $SU(2)$ is always trivially represented on states with an even weight while it acts as $-1$ on states with an odd weight. Hence $\u_i^2$ leaves $\ket{\lambda}$ invariant if $\lambda_i$ is even and sends $\ket{\lambda}$ to $\lambda(\u^2_i)\ket{\lambda}=-\ket{\lambda}$ if $\lambda_i$ is odd.

Representations of $SU(n)/Z_n$ are precisely the representations of $SU(n)$ on which the centre $Z_n$ acts trivially. On such representations, the elements of $D_{el}$ automatically act modulo the $\mathbb Z_2$ subgroup by which we factor, for even $n$, to obtain  $D_{mag}$ according to \eqref{magskelsun}. Thus the above discussion for electric representations  contains the corresponding magnetic discussion in the set of representations on which $Z_n$ acts trivially. 

Turning to the full skeleton group, we would like illustrate how one solves the constraint
\eqref{eqn:gammaconstraint}  in the case at hand. 
For any given  orbit $[\lambda,g]$ we can solve \eqref{eqn:gammaconstraint} by determining the $N_{\lambda,g}\subset W$  and choosing a representation of $N_{\lambda,g}$ which assures that the elements $(\u_i^2,(\u^*_i)^2,\u_i^2,(\u_i^*)^2)$  act trivially on the vectors $\ket{\lambda,g,v}$. If the centraliser of $[\lambda,g]$ in $\mathcal W$ is trivial its centraliser $C_{(\lambda,g)}$ in $W$ equals $D= \mathbb Z_2^{n-1}\times  \mathbb Z_2^{n-2} $. An irreducible representation $\gamma$ of $D$ is $1$-dimensional and satisfies 
$\gamma(\u_i^2,(\u_i^*)^2)=\pm 1$. The centraliser representations that satisfy the constraint \eqref{eqn:gammaconstraint} are defined by $\gamma(\u_i^2,(\u_i^*)^2)=\lambda(\u_i^2,(\u_i^*)^2)$. If $(\lambda,g) = (0,0)$ the centraliser is $W$. In this case an allowed centraliser representation $\gamma$ satisfies  $\gamma(d)\ket{v}=\ket{v}$, i.e.,~$\gamma$ is a representation of $W/D=\mathcal W$. The irreducible representations $\Pi^{[0,0]}_\gamma$  of $S$ thus correspond to irreducible representations of the permutation group $\mathcal S_n$. If $C_{(\lambda,g)}$ is neither $D$ nor $W$ the situation is more complicated and needs to be considered on a case-by-case basis for each value of $n$. 

\addcontentsline{toc}{section}{References}

\bibliography{skeletonbib}
\bibliographystyle{JHEP} %
\end{document}